\documentclass{nature_fig}
\usepackage{graphicx}
\usepackage{amsmath}

\title{Waveforms of molecular oscillations reveal circadian timekeeping mechanisms}

\author{Hang-Hyun Jo$^{1,2,3}$, Yeon Jeong Kim$^4$, Jae Kyoung Kim$^5$, Mathias Foo$^6$, \\David E. Somers$^4$ \& Pan-Jun Kim$^{7,8,9,10,*}$}

\begin{document}

\maketitle

\begin{affiliations}
\item Asia Pacific Center for Theoretical Physics, Pohang, Gyeongbuk 37673, Republic of Korea
\item Department of Physics, Pohang University of Science and Technology, Pohang, Gyeongbuk 37673, Republic of Korea
\item Department of Computer Science, Aalto University, Espoo FI-00076, Finland
\item Department of Molecular Genetics, The Ohio State University, Columbus, Ohio 43210, United States of America
\item Department of Mathematical Sciences, Korea Advanced Institute of Science and Technology, Daejeon 34141, Republic of Korea
\item School of Mechanical, Aerospace and Automotive Engineering, Coventry University, Coventry CV1 5FB, United Kingdom
\item Department of Biology, Hong Kong Baptist University, Kowloon, Hong Kong
\item Center for Quantitative Systems Biology, Hong Kong Baptist University, Kowloon, Hong Kong
\item Institute of Computational and Theoretical Studies, Hong Kong Baptist University, Kowloon, Hong Kong
\item Abdus Salam International Centre for Theoretical Physics, 34151 Trieste, Italy
\item [*]\ Correspondence and requests for materials should be addressed to P.-J.K. (email:\\ panjunkim@hkbu.edu.hk).
\end{affiliations}

\newpage

\begin{abstract}
    Circadian clocks play a pivotal role in orchestrating numerous physiological and developmental events. Waveform shapes of the oscillations of protein abundances can be informative about the underlying biochemical processes of circadian clocks. We derive a mathematical framework where waveforms do reveal hidden biochemical mechanisms of circadian timekeeping. We find that the cost of synthesizing proteins with particular waveforms can be substantially reduced by rhythmic protein half-lives over time, as supported by previous plant and mammalian data, as well as our own seedling experiment. We also find that previously-enigmatic, cyclic expression of positive arm components within the mammalian and insect clocks allows both a broad range of peak time differences between protein waveforms and the symmetries of the waveforms about the peak times. Such various peak-time differences may facilitate tissue-specific or developmental stage-specific multicellular processes. Our waveform-guided approach can be extended to various biological oscillators, including cell-cycle and synthetic genetic oscillators.
\end{abstract}

\section*{Introduction}

A variety of light-sensing organisms feature circadian clocks, which generate
endogenous molecular oscillations with $\sim$$24$~hour periodicity and thereby control numerous physiological and behavioral events\cite{Nagel2012Complexity, Sehgal1994Loss, Brody1973Circadian, Gachon2004Mammalian}. Despite the identification of biochemical mechanisms of circadian timekeeping in various organisms\cite{Fei2018Design, Hsu2014Wheels, Hurley2015Dissecting, MendozaViveros2017Molecular, Sancar2015Dawn}, our understanding of a design principle of these clock mechanisms is far from complete. For example, the mammalian clock protein BMAL1 exhibits the abundance oscillations\cite{Lee2001Posttranslational}, but these oscillations are not empirically required for the generation of circadian rhythms \emph{per se}, leaving their biological roles still unclear\cite{Liu2008Redundant, Preitner2002Orphan, McDearmon2006Dissecting}. As another example, the plant circadian system involves post-translational regulations such as the degradation of PSEUDO RESPONSE REGULATOR 5 (PRR5) protein by ZEITLUPE (ZTL) protein\cite{Kiba2007Targeted, Fujiwara2008Posttranslational}; however, a previous mathematical modeling suggests that such post-translational interactions may not be strictly required for the formation of the rhythms of any core clock components\cite{Foo2016Kernel}, raising a question about the fundamental role of these interactions.

The temporal trajectory of mRNA or protein concentration exhibiting a circadian rhythm can be characterized by its shape or waveform. A waveform of a protein expression profile, apart from its few characteristic quantities (period, amplitude, and peak phase)\cite{Hatakeyama2015Reciprocity, Luck2014Rhythmic, Patton2016Combined, Korencic2014Timing, Yoo2013Competing}, has long been underappreciated, but recently recognized for its potential importance to clock function\cite{Foo2016Kernel, Kim2012Existence, deMontaigu2015Natural}. A cuspidate waveform, which shows a notable acuteness around its peak phase, was speculated to confer high-resolution timing of downstream biological events around the peak phase\cite{Foo2016Kernel}. In addition, according to plant-clock experiments, precise changes in the waveform of \emph{GIGANTEA} (\emph{GI}) expression were sufficient to alter hypocotyl growth as a downstream phenotype\cite{deMontaigu2015Natural}. Moreover, a specific circadian waveform seems crucial for the molecular arithmetic processes involved in daily starch degradation\cite{Scialdone2013Arabidopsis}. Although not in the circadian context, there are interesting reports that modifying the waveform shape of neuro-stimulating signals changes the efficiency of entraining the neural spiking activities\cite{Cole2017Brain}. Nevertheless, the reverse yet complementary view of the waveforms as a window to the inner biochemical mechanisms of circadian clocks has not yet been taken into consideration for systematic investigation.

Here, we report that the waveforms of clock protein profiles can serve as an information source of previously-underexplored, biochemical mechanisms of circadian timekeeping. These mechanisms can be exemplified by the above PRR5-ZTL interaction and BMAL1 abundance oscillation. Interestingly, our waveform analysis predicts the considerable benefit of rhythmic regulation of protein degradation in reducing the biosynthetic cost of the waveform formation. Our mathematical framework is supported by previous, as well as our new, experimental data. This study can be extended to time-course data from various biological oscillators such as cell cycle systems and synthetic genetic oscillators.

\section*{Results}

\subsection{Relationship between waveforms and cost} 

In a circadian system, the dynamics of protein production governs the protein concentration profile $x(t)$ over time and thereby its waveform. This dynamics can often be described by the following equation:
\begin{equation}
    \frac{dx(t)}{dt} = g(t) - r(t)x(t),
    \label{eq:eq_plant}
\end{equation}
where $g(t)$ and $r(t)$ denote protein synthesis and degradation rates, respectively, as depicted in Fig.~1a. $g(t)$ is proportional to an mRNA concentration and a translation rate. Protein degradation with a rate $r(t)$ is driven by post-translational mechanisms. An oscillatory waveform of $x(t)$ satisfies $x(t)=x(t+T)$ with $T=24$~h in diurnal light and dark cycles or $T\approx 24$~h in constant light or darkness. We stress that to maintain $x(t)$'s rhythmicity, $g(t)$ or $r(t)$ should not remain constant but change over time. We will consider the relationships between $x(t)$, $g(t)$, $r(t)$, and later, the cost $c$ of protein production defined as
\begin{equation}
    c\equiv \frac{\Delta x}{T} = \langle g(t)\rangle =\langle r(t)x(t)\rangle, 
    \label{eq:cost_define}
\end{equation}
where $\Delta x$ denotes the amount of proteins synthesized over the period $T$, and $\langle \cdot \rangle$ represents a time average, e.g., $\langle g(t)\rangle\equiv (1/T)\int_0^T g(t)dt$. The equalities $\Delta x/T=\langle g(t)\rangle$ and $\langle g(t)\rangle =\langle r(t)x(t)\rangle$ are derived from Eq.~\eqref{eq:eq_plant} and $x(t)=x(t+T)$. In other words, the cost $c$ is defined as an average protein amount synthesized per time, which is equal to an average protein amount degraded per time. Because the circadian protein levels are periodic over time, the proteins must be synthesized as much as they are degraded. We will show step by step that the biosynthetic cost $c$ of a protein waveform helps us decipher circadian degradation mechanisms, mainly through the examples from the plant circadian system. Then, we will focus on other cases such as the mammalian system.

In the case of the plant \emph{Arabidopsis thaliana}, more than $20$ clock genes have been discovered, and many of their mRNAs undergo high-amplitude cycling in their abundance\cite{Nakamichi2010PseudoResponse, Flis2015Defining}. This mRNA-level oscillation is a result of transcriptional control by other clock gene products or by light signals. In the core plant clock, the protein synthesis rate $g(t)$, which is largely proportional to the transcript concentration, would likely exhibit similar oscillatory patterns. On the other hand, the characteristics of the degradation rate $r(t)$ remain rather elusive for plant clock proteins, with only a limited number of experimental reports\cite{Farre2007PRR7, Baudry2010FBox, Wang2010PRR5, Kim2003Circadian, Mas2003Targeted}. Given the clearly time-dependent nature of the protein synthesis rate, the degradation rate may not have to be also time-dependent, as demonstrated by the previous mathematical modeling\cite{Foo2016Kernel}. Existing experimental data, nonetheless, indicate that plant clock proteins often seem to have time- or phase-specific degradation rates\cite{Farre2007PRR7, Baudry2010FBox, Wang2010PRR5, Kim2003Circadian}, raising a question about the beneficial effect of such rhythmic regulation of protein stability. One study suggests that rhythmic degradation rates allow nontrivial phase differences between transcript and protein profiles\cite{Luck2014Rhythmic}. However, given that the phases of transcript profiles have relatively little functional significance, this previous study is unlikely to be about biologically beneficial effects of the rhythmic degradation rates.

We begin with the following observation: because $x(t)\geq 0$, $r(t)\geq 0$, and $g(t)\geq 0$, Eq.~\eqref{eq:eq_plant} leads to
\begin{equation}
    r(t)\geq R(t)\equiv \max\left\{-\frac{x'(t)}{x(t)}, 0\right\}.
    \label{eq:rt_condition}
\end{equation}
Note that the above inequality is always satisfied with arbitrary $g(t)\geq 0$. In other words, regardless of any specific form of a transcript profile, the protein waveform $x(t)$ imposes a stringent constraint on the protein degradation rate $r(t)$, through a lower bound $R(t)$ in Eq.~\eqref{eq:rt_condition}. Therefore, a protein waveform itself can be informative about the degradation rate.

Can waveforms indicate the effect of time- or phase-specific degradation rates observed in empirical data? In order to address this issue, we start with a contradictory scenario that the degradation rate $r(t)$ is constant over time, i.e., $r(t)=r$, and examine its consequence. From Eq.~\eqref{eq:rt_condition},
\begin{equation}
    r\geq r_{\rm min}\equiv \max_t R(t).
    \label{eq:rmin}
\end{equation}
Here, $r_{\rm min}$, the strict lower bound of the degradation rate $r$, is essentially determined only at a single time point $t = t_R$ with $t_R\equiv \arg\max_t R(t)$ ($0< t_R \leq T$; throughout this work, time $t$ in a periodic function $f(t)=f(t+T)$ is represented by a unique value within the range $0< t \leq T$, unless specified). Because $R(t) \equiv \max\{-x'(t)/x(t),0\}$, $t_R$ in practice would be a point that approaches the trough of $x(t)$ after the $x(t)$'s steepest decline ($t_R$ is placed between $t_{\rm a}$ and $t_{\rm b}$, where $t_{\rm a} \equiv \arg\max_t \{-x'(t)\}$ and $t_{\rm b} \equiv \arg\min_t x(t)$, as shown in Fig.~1b). It is surprising that only such a single time point, which will be henceforth referred to as a single hotspot, plays a critical role in determining a range of the constant degradation rate $r$. Typically, the sharper a waveform $x(t)$ is, the larger is $r_{\rm min}$ at the hotspot (Fig.~1b). 

For each plant clock protein, we can calculate the lower bound of its degradation rate, $r_{\rm min}$. Figs.~2a and 3a exhibit the empirical PRR7 and PRR5 protein profiles in equal length light-dark (12L:12D) cycles\cite{Nakamichi2010PseudoResponse}. Here, time points in light-dark cycles are counted from dawn (zeitgeber time). Using each protein profile $x(t)$, we obtain $R(t)$ in Eq.~\eqref{eq:rt_condition}, and then by Eq.~\eqref{eq:rmin}, $r_{\rm min}\approx 0.88$~h$^{-1}$ for PRR7 ($t_R\approx 21$~h) and $r_{\rm min}\approx 1.69$~h$^{-1}$ for PRR5 ($t_R\approx 22.3$~h), as in Figs.~2b and 3b. It means that if the degradation rates are constant over time, the PRR7 and PRR5 half-lives at any given time points cannot be longer than $\sim$$47$~min and $\sim$$25$~min, respectively. Provided that there are some erroneous data points in the experimental profiles, the PRR7 and PRR5 half-lives might be up to $\sim$$13$~min and $\sim$$51$~min longer than the above, respectively (Methods). In any cases, these half-lives appear to be very short, compared to other documented protein half-lives\cite{Schwanhausser2011Global, Christiano2014Global}. As previously mentioned, such a large degradation rate over the entire course of a day is attributed to only a single hotspot $t = t_R$, under the assumption that the degradation rate is constant over time. 

Next, we demonstrate that such a constant and large degradation rate can incur too large a cost of the protein production. In Eq.~\eqref{eq:cost_define}, the cost $c$ of protein production is defined as an average protein amount synthesized per time, which is equal to an average protein amount degraded per time. For a constant degradation rate $r(t)=r$, one obtains from Eqs.~\eqref{eq:cost_define} and~\eqref{eq:rmin}
\begin{equation}
    c= r\langle x(t)\rangle \geq c_{\rm g} \equiv r_{\rm min}\langle x(t)\rangle.
    \label{eq:global_cost_define}
\end{equation}
Therefore, given the protein profile $x(t)$, the lower bound of the cost $c$ (i.e., $c_{\rm g}$) is directly proportional to $r_{\rm min}$. PRR7 or PRR5, which exhibits large $r_{\rm min}$, would pay an accordingly high production cost if the degradation rate is constant. More specifically, from $cT = \Delta x \geq r_{\rm min}T \langle x(t)\rangle$, PRR7 and PRR5 must be synthesized per day at least $\sim$$21$ and $\sim$$41$ times more than actual protein level $\langle x(t)\rangle$s, respectively. In other words, these protein syntheses are far excessive compared to the actual protein abundance levels.

\subsection{Time-dependent degradation rates and cost reduction} 

The above excessive cost of protein production can be effectively alleviated by time-varying degradation rates. If the degradation rate $r(t)$ is no longer constant, $r(t)$ at $t\neq t_R$ is allowed to be smaller than $r_{\rm min}$, as far as Eq.~\eqref{eq:rt_condition} is satisfied. This fact leads to the possibility that the cost $c$ can be lower than $c_{\rm g} = r_{\rm min}\langle x(t)\rangle$. Hence, the cost can be reduced below the case of a constant degradation rate. A time-dependent degradation rate is enabled in nature by rhythmic post-translational regulation, such as PRR5 degradation by ZTL in the plant clock. Both PRR5 and ZTL levels oscillate over time, and this ZTL oscillation possibly contributes to the rhythmic degradation rate of PRR5. Including PRR5, plant clock proteins often seem to have phase-specific half-lives. These experimental data allow us to evaluate our hypothesis that rhythmic degradation rates help reduce protein production costs. 

Before the calculation of the protein production costs to examine our hypothesis, we stress that all experimental degradation rates of the plant PRR7 and PRR5 proteins and of the mouse PERIOD2 (PER2) protein\cite{Farre2007PRR7, Baudry2010FBox, Wang2010PRR5, Zhou2015Period2} are found to satisfy the fundamental relation $r(t)\geq R(t)$ in Eq.~\eqref{eq:rt_condition} (see Figs.~2b, 3b, and 4b). PER2 is an essential component of the mammalian clock\cite{Lee2001Posttranslational, Leloup2003Toward, Kim2012Mechanism, Zhou2015Period2, Vanselow2006Differential}, and its synthesis and turnover dynamics approximately follows Eq.~\eqref{eq:eq_plant}, thereby satisfying Eq.~\eqref{eq:rt_condition}. To further test the validity of Eq.~\eqref{eq:rt_condition}, we performed a cycloheximide (CHX) experiment and measured the PRR7 half-life at a time point that lacks preexisting half-life data (Fig.~2c and d and Methods). Again, the PRR7 half-life at this time point ($t=18$~h) from our own experiment is in good agreement with Eq.~\eqref{eq:rt_condition} (Fig.~2b and c). Integration of these new and previous experimental data offers a rough estimate of protein production costs, as in the following paragraphs. 

Calculation of protein production cost $c$ requires information on both degradation rate $r(t)$ and waveform $x(t)$ over time, as $c=\langle r(t)x(t)\rangle$ from Eq.~\eqref{eq:cost_define}. Because the degradation rate of each plant clock protein is only known for at most a few time points as presented above, we infer the rest degradation rates from those scarce experimental data. For this purpose, we interpolate and extrapolate the experimental degradation rate $r(t)$s based on the formula from Eq.~\eqref{eq:eq_plant}: $r(t)\approx [g(t)-x'(t)]/x(t)$. Here, the protein synthesis rate $g(t)$ can be written as $g(t) = k(t)g_{\rm m}(t)$, where $g_{\rm m}(t)$ is an mRNA concentration and $k(t)$ is an mRNA-to-protein translation rate. We discard the temporal variation of $k(t)$ and take an approximation $k(t)\approx k$. Note that experimental data of both protein and mRNA profiles, $x(t)$ and $g_{\rm m}(t)$, are available enough for a wide range of time in the cases of PRR7 and PRR5 (Fig.~2a and e and Fig.~3a and c). Using these data, one can estimate $k$ and therefore the protein degradation rate every time (see Methods). Accordingly, Figs.~2f and~3d show the estimated degradation rates of PRR7 and PRR5. Alternatively, considering the time-varying nature of $k(t)$ does not much affect our main results (Methods). In addition, we estimate the degradation rate $r(t)$ of PER2. Experimental degradation rates of PER2 cover a relatively wide range of time and are thus informative enough to envisage the overall trend of $r(t)$. Therefore, only based on these experimental degradation rates and $R(t)$, without mRNA profile data, we can make a rough estimate of the PER2 degradation rate over the entire circadian period, as demonstrated in Fig.~4c (Methods).

The estimated, phase-specific degradation rates of clock proteins in Figs.~2f, 3d, and 4c show the characteristic curves that peak around the hotspots ($t\approx t_R$) and decline elsewhere. These patterns are the hallmarks of the rhythmic degradation rates that can reduce the protein production costs below the cases of constant degradation rates; except for the hotspots that must have large degradation rates ($\geq r_{\rm min}$) by Eq.~\eqref{eq:rt_condition}, if degradation rates remain small for most time, proteins do not have to be much synthesized to compensate for the degradation (Eq.~\eqref{eq:cost_define}) and hence the production costs will become reduced.
 
Using the above degradation rate curves of several clock proteins, we now compute the actual protein production cost $c$ by $c=\langle r(t)x(t)\rangle$ in Eq.~\eqref{eq:cost_define}. Compared to the cases of constant degradation rates, the PRR7, PRR5, and PER2 production costs indeed decrease by at least $\sim$$70\%$, $\sim$$83\%$, and $\sim$$52\%$, respectively, as summarized in Table~1. If we consider a possible deviation of the degradation rate in Fig.~2c, the PRR7 production cost decreases by $\sim$68\% to $\sim$73\% (Fig.~2). Interestingly, in the case of alga \emph{Ostreococcus tauri}, rhythmic protein degradation is known to be very crucial for circadian timekeeping\cite{vanOoijen2011Proteasome}. For its clock proteins CIRCADIAN CLOCK ASSOCIATED 1 (CCA1) and TIMING OF CAB EXPRESSION 1 (TOC1), the full time series of experimental $r(t)$ is available\cite{vanOoijen2011Proteasome}, and our analysis suggests that the rhythmic $r(t)$ saves $\sim$30\% and $\sim$41\% of the CCA1 and TOC1 production costs, respectively (Methods). These results well support our hypothesis that rhythmic control of clock protein half-lives is beneficial to the cost reduction of protein production. This cost saving effect would be valid even if other benefits from the rhythmic half-lives are not clear. We thus predict the statistical tendency that the sharper a waveform is at the hotspot (i.e., the larger is $r_{\rm min}$, and therefore is $c_{\rm g}$), the more likely a protein half-life is to be phase-specific.

\subsection{Enigmatic elements of animal circadian systems}

Thus far, we have investigated circadian dynamics driven by protein synthesis and degradation in Eq.~\eqref{eq:eq_plant}. We now discuss another class of circadian dynamics with Eq.~\eqref{eq:eq_animal} below, which is crucial for mammals and insects, but does not follow the underlying mechanism of Eq.~\eqref{eq:eq_plant}.

The core part of the mammalian clock harbors a transcriptional/post-translational negative feedback loop\cite{Lee2001Posttranslational, Zhou2015Period2, Leloup2003Toward}, which involves transcription factors, CLOCK and BMAL1 proteins. CLOCK-BMAL1 heterodimers activate the transcription of \emph{Per} and \emph{Cryptochrome} (\emph{Cry}) genes, and the encoded PER and CRY proteins form PER-CRY complexes that are translocated to the nucleus. In the nucleus, they interact with CLOCK-BMAL1 complexes to inhibit the CLOCK-BMAL1 transcriptional activities. These positive (CLOCK and BMAL1) and negative (PER and CRY) arms constitute a negative feedback loop.

In the following equations, $x_{\rm A}(t)$ and $x_{\rm I}(t)$ represent the concentrations of active and inactive CLOCK-BMAL1 complexes in the nucleus, respectively, and $y(t)$ represents the concentration of nuclear PER-CRY complexes that are not binding to CLOCK-BMAL1 complexes:
\begin{eqnarray}
    \frac{dx_{\rm A}(t)}{dt} &=& \tilde\alpha(t) +k_1x_{\rm I}(t)-ky(t)x_{\rm A}(t) -r_1x_{\rm A}(t),\\
    \frac{dx_{\rm I}(t)}{dt} &=& ky(t)x_{\rm A}(t) -k_1x_{\rm I}(t) -r_2x_{\rm I}(t).
\end{eqnarray}
Here, $\tilde\alpha(t)$ is a rate of CLOCK-BMAL1 translocation from the cytoplasm to the nucleus, $k_1$ and $k$ are, respectively, dissociation and association rate constants of two complexes, CLOCK-BMAL1 and PER-CRY, and $r_1$ and $r_2$ correspond to the sums of degradation rates and the rates of translocation to the cytoplasm. Employing another variable $x_{\rm n}(t)\equiv x_{\rm A}(t)+x_{\rm I}(t)$ for the total CLOCK-BMAL1 concentration, the upper equation can be rewritten as
\begin{equation}
    \frac{dx_{\rm A}(t)}{dt} = g_{\rm A}(t) - [r_0+ky(t)] x_{\rm A}(t),
    \label{eq:eq_animal}
\end{equation}
where $g_{\rm A}(t) \equiv k_1x_{\rm n}(t)+\tilde\alpha(t)$ and $r_0\equiv k_1+r_1$.

Equation~\eqref{eq:eq_animal} represents a class of circadian dynamics distinguished from our previous case, Eq.~\eqref{eq:eq_plant}. Equation~\eqref{eq:eq_animal} for the core mammalian clock captures the dynamics of active CLOCK-BMAL1 complexes (i.e., CLOCK-BMAL1 that is not binding to PER-CRY) in the nucleus, as depicted in Fig.~5a, and is applied to the insect clock as well. A fundamental difference between Eq.~\eqref{eq:eq_plant} and Eq.~\eqref{eq:eq_animal} is as follows: in Eq.~\eqref{eq:eq_plant}, $g(t)$ exhibits high-amplitude oscillation as evident from the transcript profiles of many plant clock genes and mammalian \emph{Per} genes, and hence $g(t)$ is a main driving force of $x(t)$'s oscillation. In contrast, in Eq.~\eqref{eq:eq_animal}, $x_{\rm A}(t)$'s oscillation is largely driven by $y(t)$'s oscillation, rather than by $g_{\rm A}(t)$'s. Compared to PER2 levels ($\propto y(t)$; Fig.~4a), BMAL1 levels are only weakly oscillating over time\cite{Lee2001Posttranslational}, and correspondingly, $g_{\rm A}(t)$ would be only weakly oscillating. In fact, cyclic BMAL1 expression is not even required for mammalian circadian rhythmicity, as the mutant with constitutive BMAL1 expression still exhibits circadian rhythms\cite{Liu2008Redundant, Preitner2002Orphan, McDearmon2006Dissecting}.

Given the apparently minor role of BMAL1's abundance oscillation in circadian rhythmicity, what beneficial effects on the clock might follow from this BMAL1 oscillation?

\subsection{Diverse phase differences between clock components}

To the above enigmatic presence of BMAL1's abundance oscillation in the mammalian clock, the effect of protein production cost $c$ in our previous analysis is not straightforwardly relevant. Unlike $x(t)$ in Eq.~\eqref{eq:eq_plant}, $x_{\rm A}(t)$ involves the only active, not the total, molecules. In other words, $x_{\rm A}(t)$ is mainly driven by the relatively costless, post-translational conversion of inactive to active molecular forms, devoid of severe biosynthetic cost problems in the previous analysis. Again, we suggest that the clue for the effect of the BMAL1 oscillation can be found from waveforms, especially $x_{\rm A}(t)$ and $y(t)$ from the CLOCK-BMAL1 and PER-CRY complexes. As will be shown later, such BMAL1 oscillation confers at least two advantages on the circadian system: one is a wide range of a peak time difference between the two clock components, active CLOCK-BMAL1, and PER-CRY that is not binding to CLOCK-BMAL1. The other advantage is the symmetry of the waveforms of these components. For the sake of convenience, we will henceforth drop subscript As from $x_{\rm A}(t)$ and $x_{\rm A}'(t)$, and simply write them as $x(t)$ and $x'(t)$. Equation~\eqref{eq:eq_animal} can be rewritten as
\begin{equation}
    ky(t)=\frac{g_{\rm A}(t)-x'(t)}{x(t)} - r_0.
    \label{eq:eq_animal_y}
\end{equation}
If we assume that $g_{\rm A}(t)$ ($\propto$ BMAL1 level) is completely constant over time, i.e., $g_{\rm A}(t)=g$, waveforms $y(t)$ and $x(t)$ in Eq.~\eqref{eq:eq_animal_y} are substantially constrained by each other. Specifically,
\begin{eqnarray}
    \label{eq:const_g_y}
    ky(t) &=& \frac{g - x'(t)}{x(t)} - r_0,\\
    g &\geq & g_{\rm min}\equiv \max_t x'(t),
    \label{eq:g_min}
\end{eqnarray}
where the inequality of the lower relation comes from $x(t)\geq 0$ and $r_0+ky(t)\geq 0$. We will use a notation $t_f\equiv \arg\max_t f(t)$ ($0< t_f\leq T$) for any given periodic function $f(t)$ ($f(t)=f(t+T)$) when $t_f$ is uniquely determined by $0< t_f \leq T$. For example, $t_x$ denotes the peak time of $x(t)$ during $0< t \leq T$. From Eq.~\eqref{eq:const_g_y},
\begin{equation}
    t_{-\frac{x'}{x}} \leq t_y\leq t_{\frac{1}{x}}.
    \label{eq:t_y_t_R}
\end{equation}
Note that $t_{-\frac{x'}{x}}$ is identical to the hotspot $t_R$. In other words, $t_y$ is even closer to $x(t)$'s trough time than $t_R$. This range of $t_y$ is illustrated in Fig.~5b. Generally, a peak time difference between $y(t)$ and $x(t)$ takes such a narrow range that $y(t)$'s peak time is almost the same as $x(t)$'s trough time. Therefore, if $g_{\rm A}(t)$ stays constant over time, waveforms $x(t)$ and $y(t)$ are only allowed to have a near anti-phase relationship.

To exemplify the above point, we consider the case with a sinusoidal wave $x'(t)=L\sin(\omega t)$ where $L$ is a constant and $\omega\equiv 2\pi/T$. In this case,
\begin{equation}
    x(t)= -\frac{L}{\omega} \cos(\omega t) +\frac{L}{\omega}+h_0,
    \label{eq:sine_wave}
\end{equation}
with an additional constant $h_0$. In the subsequent analyses, we treat $x(t)$ in Eq.~\eqref{eq:sine_wave} as dimensionless, without loss of generality. We define a phase difference between $x(t)$ and $y(t)$ in Eq.~\eqref{eq:const_g_y} as $\phi \equiv |\omega(t_y-t_x)|$. Using Eq.~\eqref{eq:const_g_y}, $\phi=\pi - 2\tan^{-1}[(\sqrt{C^2-L^2+g^2} -g)/(C+L)]$, where $C\equiv h_0\omega +L$ and $g\geq g_{\rm min}=L$ from Eq.~\eqref{eq:g_min}. This exact solution of $\phi$ is indeed very close to $\pi$, as shown in Fig.~5c. This tendency is consistent with our generic result that constant $g_{\rm A}(t)$ forces $x(t)$ and $y(t)$ into a near anti-phase relationship.

In contrast, if $g_{\rm A}(t)$ is no longer constant but cycles over time, as observed with cyclic BMAL1 expression in nature, then waveforms $y(t)$ and $x(t)$ in Eq.~\eqref{eq:eq_animal_y} are not much constrained by each other, and their peak time difference (or phase difference) can be flexible depending on $g_{\rm A}(t)$'s oscillatory form. Because there is a lack of compelling experimental data on the waveform of $g_{\rm A}(t)$, we start with the following assumption:
\begin{equation}
    g_{\rm A}(t) \approx \alpha + \beta x(t+\tau),
    \label{eq:mammal_g_assume}
\end{equation}
where $\alpha$, $\beta$, and $\tau$ are constants, and $\beta\geq 0$ and $\tau\geq 0$. From Eq.~\eqref{eq:eq_animal_y}, $r_0+ky(t)\geq 0$, and $g_{\rm A}(t)\geq 0$, $\alpha$ should satisfy
\begin{equation}
    \alpha \geq \alpha_{\rm min}\equiv \max\left\{ \max_t [x'(t) -\beta x(t+\tau)], \max_t [-\beta x(t)]\right\}.
    \label{eq:mammal_alpha_min}
\end{equation}
Next, we show that $x(t)$ and $y(t)$ can have almost any in-phase to anti-phase relationship, covering a wide range of the phase difference. If $\tau\ll T$,
\begin{equation}
    x(t+\tau)\approx x(t)+\tau x'(t).
    \label{eq:x_t_tau_expansion}
\end{equation}
Combined with Eq.~\eqref{eq:eq_animal_y}, it leads to
\begin{equation}
    ky(t) \approx \frac{\alpha - (1-\beta\tau)x'(t)}{x(t)} +\beta-r_0.
    \label{eq:varying_g_y}
\end{equation}
Depending on signs of $\alpha$ and $1-\beta\tau$ in Eq.~\eqref{eq:varying_g_y}, $y(t)$ is now allowed to peak anytime of a day relative to $x(t)$'s peak time, as proven in Methods.

This result is illustrated in Fig.~5d. Together, if BMAL1 level ($\propto g_{\rm A}(t)$ in Eq.~\eqref{eq:eq_animal_y} and Eq.~\eqref{eq:eq_animal}) is not constant but varies over time, it confers much freedom on the waveform $y(t)$ of PER-CRY that is not binding to CLOCK-BMAL1, and thus allows various phase differences between those unbinding CLOCK-BMAL1 and PER-CRY complexes ($x(t)$ and $y(t)$) through the adjustment of parameters $\alpha$, $\beta$, and $\tau$. As in Fig.~5d, the unbinding CLOCK-BMAL1 and PER-CRY complexes can take almost any in-phase to anti-phase relationship. This result is in sharp contrast to the case with constant $g_{\rm A}(t)$, where the unbinding CLOCK-BMAL1 and PER-CRY complexes have a predominantly anti-phase-like relationship. Our predictions can be verified by experimental techniques, such as co-immunoprecipitation assays, measuring the time series of CLOCK-BMAL1 and PER-CRY levels across different tissues or developmental stages, while excluding the levels of inactive CLOCK-BMAL1.

These potentially diverse phase differences, conferred by cyclic expression of positive arm components, may help in the coordination of tissue-specific or developmental stage-specific clock events in complex multicellular organisms, such as mammals and insects\cite{Yagita2010Development, Korencic2014Timing}. Interestingly, the fungus \emph{Neurospora crassa}, a relatively simple species, shows almost constant levels of \emph{white collar-1} (\emph{wc-1}) expression\cite{Merrow2001Circadian}, and thus would have almost constant $g_{\rm A}(t)$. Therefore, we expect that the fungal clock may have an only anti-phase-like relationship between its core components, nuclear WC-1 and FREQUENCY (FRQ) proteins (see Supplementary Discussion).

To illustrate the above diverse phase differences conferred by BMAL1 cycling, we revisit the case with a sinusoidal wave $x(t)$ in Eq.~\eqref{eq:sine_wave} and consider the oscillation of $g_{\rm A}(t)$ in Eq.~\eqref{eq:mammal_g_assume}. From Eq.~\eqref{eq:eq_animal_y}, we obtain the exact solution of the phase difference between $x(t)$ and $y(t)$, as plotted in Fig.~5e. This exact solution is in good agreement with our generic results based on the approximation Eqs.~\eqref{eq:x_t_tau_expansion} and~\eqref{eq:varying_g_y}.

\subsection{Symmetries of waveforms}

Another advantage of cyclic expression of positive arm components in mammalian and insect clocks is in the symmetry of waveforms about the peak phases. Previous experimental data from the mammalian clock indicate the existence of such symmetry that ascending and descending phases span almost the same time intervals\cite{Meng2008Setting}, while its phenotypic significance still remains unknown.

To demonstrate the effect of BMAL1 cycling on the waveform symmetry, we first assume the contradictory scenario that $g_{\rm A}(t)$ is constant over time with $g_{\rm A}(t)=g$. Therefore, $ky(t) = [g - x'(t)]/x(t) - r_0$ from Eq.~\eqref{eq:const_g_y}. In this case, waveforms $x(t)$ and $y(t)$ from CLOCK-BMAL1 and PER-CRY complexes cannot easily satisfy both symmetric relations $x(t_x-t) \approx x(t_x+t)$ and $y(t_y-t) \approx y(t_y+t)$ at the same time, because $x'(t)$ term in $ky(t) = [g - x'(t)]/x(t) - r_0$ breaks the symmetry of either $x(t)$ or $y(t)$ waveform unless $g \gg \max_t |x'(t)|$ to diminish the effect of $x'(t)$.

In contrast, if BMAL1 level ($\propto g_{\rm A}(t)$) is not constant but varies over time, both unbinding CLOCK-BMAL1 and PER-CRY profiles ($x(t)$ and $y(t)$) are allowed to have symmetric waveforms relatively easily. For example, if $g_{\rm A}(t)\approx \alpha + \beta x(t+\tau)$ and $x(t+\tau)\approx x(t)+\tau x'(t)$ with $\tau\ll T$ as in Eqs.~\eqref{eq:mammal_g_assume} and~\eqref{eq:x_t_tau_expansion}, then $ky(t)\approx [\alpha - (1-\beta\tau)x'(t)]/x(t)+\beta -r_0$ in Eq.~\eqref{eq:varying_g_y}. Therefore, both $x(t)$ and $y(t)$ waveforms can be approximately symmetric at the same time, as long as $|\alpha/(1-\beta\tau)| \gg \max_t |x'(t)|$ for the diminished effect of $x'(t)$. This condition can be satisfied more easily than the previous one.

This waveform symmetry, along with the above phase difference between two core components (unbinding CLOCK-BMAL1 and PER-CRY complexes), shows that the waveforms are useful to understand the effect of the enigmatic oscillation in BMAL1 expression.

\section*{Discussion}\label{sect:conclusion}

In this study, we have revealed that protein waveforms are informative about the underlying mechanisms of circadian clockwork.

A sharp waveform at the hotspot time point (i.e., with large $r_{\rm min}$) implies rhythmic post-translational regulation that yields a phase-specific protein half-life; otherwise, too large costs of protein syntheses can be incurred for those waveforms. Such rhythmic degradation rates are observed in plant and mammalian circadian clocks, and can substantially reduce the protein production costs, as demonstrated in Table~1. If more experimental data become available, our waveform-cost analysis can be extended to other clock proteins. For example, the orphan nuclear receptor REV-ERB$\alpha$ in the mammalian clock may have a phase-specific half-life, driven by the rhythmic activity of glycogen synthase kinase-$3\beta$ that regulates the REV-ERB$\alpha$ stability\cite{Yin2006Nuclear, Iitaka2005Role, Besing2015Circadian}. Hence, if the half-lives measured at multiple specific time points become available, REV-ERB$\alpha$ will be a good target candidate for our cost analysis, aided by the existing REV-ERB$\alpha$ expression profiles\cite{Narumi2016Mass}.

On the other hand, regarding any possible extra costs that may be incurred by rhythmic degradation rates, we note that the cost $c$ of a given protein is not the concept to include the production cost of its proteolytic factor. Yet, the half-life can exhibit a rhythmic pattern by the proteolytic factor's oscillation, and thus one may suggest that the cost $c$ should be extended to the proteolytic factor's production cost. This extra cost from the proteolytic factor production, however, is not always relevant and needs cautious analyses in the future. For example, if the proteolytic factor has not only evolved for the degradation of a particular protein but also for other functions, then the cost of the proteolytic factor production shall not be covered by the cost $c$ in question. This is because such a proteolytic factor continues to be produced for multiple purposes, not exclusively for the degradation of that particular protein. 

In this study, we also suggest that seemingly dispensable, cyclic expression of certain clock proteins in mammals and insects may allow both a broad range of phase differences between clock components and the symmetries of the waveforms. The various phase differences may be important for tissue-specific or developmental stage-specific clock coordination in complex multicellular organisms, such as mammals and insects. As previously mentioned, fungi do not show such cyclic expression of the corresponding components, and their relatively simple organismal forms may not necessitate as widely-ranging phase differences as in the cases of mammals and insects.

Our waveform-guided approach is well supported by experimental data (Figs.~2b, 3b, and 4b), and provides insights into circadian mechanisms of evolutionarily-distant organisms\cite{Hsu2014Wheels, Hurley2015Dissecting, MendozaViveros2017Molecular}. Furthermore, we envisage that the concepts presented in this study can be applied beyond circadian dynamics, such as to time-course data from cell cycle systems and synthetic genetic oscillators\cite{Chen2015Emergent, Tyson2015Models, OKeeffe2017Oscillators}.

\begin{methods}

\subsection{Experimental measurement of the PRR7 half-life}
    
We describe the details of our experimental methods for the measurement of the PRR7 degradation rate, of which data are available in Fig.~2c and d and Supplementary Figs.~1 and 2. For CHX assays, \emph{PRR7pro::FLAG-PRR7-GFP} seedlings were grown on MS media with 3\% sucrose and 1\% agar under 12L:12D cycles (white fluorescent light; $30$--$40$~$\mu$mol~m$^{-2}$~s$^{-1}$) at $22\,^{\circ}$C for $14$ days. Seedlings were transferred to MS liquid media with $100$~$\mu$M CHX or mock (ethanol) at ZT17 in darkness. The tissues were kept in the dark under slow shaking and collected at $0$, $1$, $3$, $5$, and $7$~h post treatment. 

For immunoblots, the tissue was ground in liquid nitrogen and extracted in protein extraction buffer ($50$~mM Tris-Cl, pH~$7.5$, $150$~mM NaCl, 0.5\% Nonidet P-40, $1$~mM EDTA, $3$~mM dithiothreitol, $1$~mM phenylmethylsulfonyl fluoride, $5$~$\mu$g~ml$^{-1}$ leupeptin, $1$~$\mu$g~ml$^{-1}$ aprotinin, $1$~$\mu$g~ml$^{-1}$ pepstatin, $5$~$\mu$g~ml$^{-1}$ antipain, $5$~$\mu$g~ml$^{-1}$ chymostatin, $50$~$\mu$M MG132, $50$~$\mu$M MG115, $50$~$\mu$M ALLN). Total proteins were separated using an 8\% SDS-PAGE gel (acrylamide:bisacrylamide, 37.5:1), immunoblotted and probed with anti-GFP antibody (Abcam, ab6556) and polyclonal anti-ADK antibody (gift from Dr. David Bisaro) diluted to 1:4000 and 1:15000, respectively, followed by anti-rabbit IgG conjugated with horseradish peroxidase (GE healthcare, NA934). Chemiluminescent detection was performed using SuperSignal\textsuperscript{TM} West Pico Chemiluminescent Substrate (Thermo Scientific, 34080). The FLAG-PRR7-GFP protein signals were calculated by ImageJ software (NIH, version 1.8.0) from three biological repeats, and were normalized to their corresponding ADK (adenosine kinase) signal intensities individually. 

All unique biological materials used in this study (\emph{PRR7pro::FLAG-PRR7-GFP}) are available from the authors upon request.

\subsection{Analysis of data from the plant and mammalian clocks}

By writing the protein synthesis rate $g(t)$ as $g(t) = k(t)g_{\rm m}(t)$ and by assuming the roughly constant $k(t)$, i.e., $k(t)\approx k$, Eq.~\eqref{eq:eq_plant} can be written as
\begin{equation}
    \frac{dx(t)}{dt}\approx kg_{\rm m}(t) - r(t)x(t),
    \label{eq:eq_plant_mRNA}
\end{equation}
which leads to
\begin{equation}
    k\approx \frac{x'(t_i) + r(t_i)x(t_i)}{g_{\rm m}(t_i)},
\label{eq:eq_k_infer}
\end{equation}
where $t_i$ corresponds to each time point $t=t_i$ with experimentally-available degradation rate $r(t)$. In the case of PRR7, we used the protein degradation rates at $t_i =4$~h, $12$~h, and $18$~h (Fig.~2 and Supplementary Fig.~3). The first two degradation rates were obtained from the protein abundance data in Fig.~7b of Farre et al.\cite{Farre2007PRR7}, while the last degradation rate was from our own experimental data in Fig.~2c. We also obtained the experimental data of the mRNA and protein profiles from Fig.~5d of Flis et al.\cite{Flis2015Defining} and Fig.~5a of Nakamichi et al.\cite{Nakamichi2010PseudoResponse}, respectively. Both datasets have $2$-hour sampling intervals under 12L:12D cycles. These mRNA and protein levels were normalized by the peak levels of their splines, and adopted for $g_{\rm m}(t)$ and $x(t)$ in Eq.~\eqref{eq:eq_plant_mRNA}, respectively. From Eq.~\eqref{eq:global_cost_define}, $c_{\rm g} \approx 0.40$~h$^{-1}$. Using $r(t_i)$, $x(t_i)$, and $g_{\rm m}(t_i)$, we obtained $k$ from Eq.~\eqref{eq:eq_k_infer}. To be precise, although we treat $k$ as a constant, different $t_i$s can have different $k$ values calculated from Eq.~\eqref{eq:eq_k_infer}. For simplicity of our analysis, we discarded such differences and took the average of $k$ over $t_i$. Using this $k$, we inferred $r(t)$ for the rest of time ($t\neq t_i$) by the following formula from Eq.~\eqref{eq:eq_plant_mRNA}:
\begin{equation}
    r(t)\approx \frac{k g_{\rm m}(t)-x'(t)}{x(t)}.
    \label{eq:eq_plant_mRNA_rt}
\end{equation}
Because experimental protein and mRNA levels have $2$-hour sampling intervals, we inferred degradation rate $r(t)$ every $2$ hours, except for $t=4$~h, $12$~h, and $18$~h for which we used experimentally-known $r(t)$ values. The overall $r(t)$ profile exhibits two peaks at $20$~h~$\leq t \leq 22$~h and at $2$~h~$\leq t \leq 10$~h. The former peak is a natural consequence of large $R(t)$ around that time (red solid line in Supplementary Fig.~3a), while the latter may be an artifact from unconsidered biological factors. To reduce the effect of such possible artifact, we replace every $r(t)> \max_{20\textrm{ h} \leq t \leq 22\textrm{ h}} r(t)$ by $\max_{20\textrm{ h} \leq t \leq 22 \textrm{ h}} r(t)$, because $\max_{20\textrm{ h} \leq t \leq 22\textrm{ h}} r(t)\approx 1.02$~h$^{-1}$ and the real degradation rate is unlikely to be larger than $1.02$~h$^{-1}$. We also replace every $r(t)< \min\{r(t=4$~h$), r(t=12$~h$), r(t=18$~h$)\}$ by $\min\{r(t=4$~h$), r(t=12$~h$), r(t=18$~h$)\}$, and therefore the lower bound of $r(t)$ is set to the minimum value of experimental $r(t)$ values. In such a way, the difference between $c$ and $c_{\rm g}$ is reduced (Eqs.~\eqref{eq:cost_define} and~\eqref{eq:global_cost_define}), leading to a conservative estimate of that difference. The resulting $r(t)$ is presented in Supplementary Fig.~3a. Because $r(t)$ at $2$~h~$\leq t \leq 10$~h is improbably deviated from the overall trend of experimental $r(t)$ values, we correct this part by linear interpolation and extrapolation of the experimental $r(t=4$~h$)$ and $r(t=12$~h$)$ values, as shown in Fig.~2f. Consequently, $c\approx 0.30 c_{\rm g}$ with $r(t)$ in Fig.~2f and $c\approx 0.67c_{\rm g}$ with $r(t)$ in Supplementary Fig.~3a. In other words, whether correcting $r(t)$ at $2$~h~$\leq t \leq 10$~h or not, the actual cost of PRR7 waveform maintenance would be at most one- to two-thirds of the assumed cost in the case of a constant degradation rate.

Thus far, we have adopted the experimental protein levels for $x(t)$. However, we suppose that experimental protein levels, when low around a trough phase, can be susceptible to measurement errors. Such potentially inaccurate data, if these data underestimate the protein levels around the trough phase, can lead to the overestimation of $r_{\rm min}$ in Eq.~\eqref{eq:rmin} and $c_{\rm g}$ in Eq.~\eqref{eq:global_cost_define}, and thereby exaggerate a difference between $c_{\rm g}$ and $c$. To mitigate these possibly erroneous effects, we consider a new $x(t)$ whose values at $t=0$~h, $22$~h, and $24$~h are replaced by that of $x(t=2$~h$)$, as plotted in Supplementary Fig.~3b. With this smoothened $x(t)$, we obtain $r_{\rm min} \approx 0.69$~h$^{-1}$, which is smaller than $r_{\rm min}\approx 0.88$~h$^{-1}$ from the original $x(t)$. Likewise, new $c_{\rm g} \approx 0.32$~h$^{-1}$ and $c\approx 0.12$~h$^{-1}$. Here, $c$ is calculated from the newly estimated $r(t)$ in Supplementary Fig.~3c. On the other hand, without a correction for $2$~h~$\leq t \leq 10$~h as in Supplementary Fig.~3d, $c\approx 0.22$~h$^{-1}$. Still, the cost of PRR7 waveform maintenance is at most one- to two-thirds of the assumed cost in the case of a constant degradation rate. These results are summarized in Supplementary Table~1.

In the case of PRR5, we used experimental protein degradation rates at $t_i =12$~h and $19$~h (Fig.~3 and Supplementary Fig.~4) from the protein abundance data in Fig.~7c of Baudry et al.\cite{Baudry2010FBox}. We obtained the experimental data of the mRNA and protein profiles from Fig.~5d of Flis et al.\cite{Flis2015Defining} and Fig.~5a of Nakamichi et al.\cite{Nakamichi2010PseudoResponse}, respectively. Both datasets have $2$-hour sampling intervals under 12L:12D cycles. These mRNA and protein levels were normalized by the peak levels of their splines, and adopted for $g_{\rm m}(t)$ and $x(t)$ in Eq.~\eqref{eq:eq_plant_mRNA}, respectively. Following a similar procedure to the case with PRR7, we obtained $c_{\rm g} \approx 0.77$~h$^{-1}$, and inferred the degradation rate $r(t)$ every $2$ hours, except for $t=12$~h and $19$~h for which we used experimentally-known $r(t)$ values. When calculating $c$ based on this inferred $r(t)$, we replace every $r(t)> r_{\rm min}$ by $r_{\rm min}$, because the real degradation rate is unlikely to be larger than $r_{\rm min}\approx 1.69$~h$^{-1}$. We also replace every $r(t)< \min\{r(t=12$~h$), r(t=19$~h$)\}$ by $\min\{r(t=12$~h$), r(t=19$~h$)\}$, and therefore the lower bound of $r(t)$ is set to the minimum value of experimental $r(t)$ values. In such a way, the difference between $c$ and $c_{\rm g}$ is reduced, leading to a conservative estimate of that difference. The resulting $r(t)$ is presented in Supplementary Fig.~4a. Because $r(t)$ at $6$~h~$\leq t \leq 10$~h is improbably deviated from the overall trend of experimental $r(t)$ values, we correct this part by linear extrapolation of the experimental $r(t=12$~h$)$ value, as shown in Fig.~3d. Consequently, $c\approx 0.17 c_{\rm g}$ with $r(t)$ in Fig.~3d and $c\approx 0.34 c_{\rm g}$ with $r(t)$ in Supplementary Fig.~4a. In other words, whether correcting $r(t)$ at $6$~h~$\leq t \leq 10$~h or not, the actual cost of PRR5 waveform maintenance would be at most one-sixth to one-third of the assumed cost in the case of a constant degradation rate.

To mitigate the aforementioned, possibly erroneous effects from low protein levels around a trough phase, we consider new $x(t)$ whose values at $t=0$~h, $22$~h, and $24$~h are increased as in Supplementary Fig.~4b. With this smoothened $x(t)$, we obtain $r_{\rm min} \approx 0.55$~h$^{-1}$, which is smaller than $r_{\rm min}\approx 1.69$~h$^{-1}$ from the original $x(t)$. Likewise, new $c_{\rm g} \approx 0.26$~h$^{-1}$ and $c\approx 0.13$~h$^{-1}$. Here, $c$ is calculated from the newly estimated $r(t)$ in Supplementary Fig.~4c. On the other hand, without a correction for $6$~h~$\leq t \leq 10$~h as in Supplementary Fig.~4d, $c\approx 0.17$~h$^{-1}$. Still, the cost of PRR5 waveform maintenance is at most one-half to two-thirds of the assumed cost in the case of a constant degradation rate. These results are summarized in Supplementary Table~1.

For PRR7 and PRR5, the time-varying nature of $k(t)$ in $g(t) = k(t)g_{\rm m}(t)$ can be considered as an alternative to the above possibility $k(t) \approx k$. Because of a lack of data on the genuine form of $k(t)$ for these proteins, we tried a sinusoidal approximation $k(t) \approx \max\{a \sin(2\pi t/T-\phi)+b, \epsilon_k\}$, where $a$, $b$, and $\phi$ are constants that fit the function $a \sin(2\pi t/T-\phi)+b$ to the right-hand side of Eq.~\eqref{eq:eq_k_infer} and $\epsilon_k$ is a small positive constant to ensure $k(t) > 0$ ($\epsilon_k$ was set to the minimum value of the right-hand side of Eq.~\eqref{eq:eq_k_infer}). In the PRR5 case, $a$, $b$, and $\phi$ were underdetermined, and thus $a$ and $b$ were obtained for each value of $\phi$ in the range $0\leq \phi\leq \pi/2$ (which does not involve any loss of generality for the PRR5 data). Applying such $k(t)$ to Eq.~\eqref{eq:eq_plant_mRNA_rt}, instead of $k$ therein, and repeating all the above procedures (Supplementary Fig.~5) did not much change our results: under this assumption of the time-varying $k(t)$, the estimated rhythmic degradation rates led to $\sim$73\% reduction of the PRR7 production cost and $\sim$83$\sim$84\% reduction of the PRR5 production cost (cf., Table~1 for the case $k(t) \approx k$).

In the case of the mouse PER2 protein, we obtained the time-course abundance data of the CHX-untreated control in Fig.~1a of Zhou et al.\cite{Zhou2015Period2}, and adopted this protein profile for $x(t)$. The original profile covers $\sim$$45$-hour-long data with $0.1$-hour resolution. Therefore, we considered the data at $9.6$~h~$\leq t \leq 33$~h for one circadian period ($T=23.4$~h), and smoothened them with a moving window average ($3$-hour window). These data were normalized by their peak level, and the resulting $x(t)$ appears in Fig.~4a. $R(t)$ derived from this $x(t)$ is very noisy, and therefore smoothened with a moving window average ($1$-hour window). For experimental protein degradation rates, we used the instantaneous half-lives after $0.5$ hours since CHX treatment at $t=19$~h, $22$~h, $25$~h, $28$~h, and $30$~h in Supplementary Fig.~1a of Zhou et al.\cite{Zhou2015Period2}.

Full details of the PRR7, PRR5, and PER2 data collection are provided in Supplementary Methods and Supplementary Table 1.

\subsection{Analysis of data from the algal clock}

In the case of CCA1 and TOC1 proteins in the \emph{Ostreococcus} circadian system, we obtained the full time-course degradation rate $r(t)$ and protein level $x(t)$ data from Fig.~1a and~b of van Ooijen et al.\cite{vanOoijen2011Proteasome}, respectively (12L:12D-cycle condition). We did not perform any normalization of $x(t)$, and the unit of $x(t)$ here follows that of van Ooijen et al.\cite{vanOoijen2011Proteasome} (molecules$\cdot$cell$^{-1}$). Because $x(t)$'s sampling resolution was rather low (4-hour sampling interval), we did not apply $r(t)$ and $x(t)$ to Eq.~\eqref{eq:rt_condition} wherein the specific form of $R(t)$ could be sensitive to the $x(t)$'s sampling resolution. For the calculation of $c_{\rm g}$, we estimated $r_{\rm min}$ as $r_{\rm min} \approx \min\{\max_t r(t), \max_t R(t)\}$, with regards to possibly-inaccurate $R(t)$ from the low sampling resolution of $x(t)$. For the calculation of $c$, we adopted $r(t)x(t)$ in Fig.~1c of van Ooijen et al.\cite{vanOoijen2011Proteasome}. As a result, for CCA1 and TOC1, $r_{\rm min}\approx 0.25$ and $0.28$~h$^{-1}$, $c_{\rm g} \approx 60.7$ and $19.7$ molecules$\cdot$cell$^{-1}\cdot$h$^{-1}$, and $c \approx 42.5$ and $11.6$ molecules$\cdot$cell$^{-1}\cdot$h$^{-1}$, respectively. In other words, the cost of CCA1 and TOC1 production is about two-thirds of the assumed cost in the case of constant degradation rates.

\subsection{Effects of oscillating BMAL1 expression}

If $\tau\ll T$ in Eq.~\eqref{eq:mammal_g_assume}, Eq.~\eqref{eq:varying_g_y} can be used to calculate a phase difference between $x(t)$ and $y(t)$. Without loss of generality, let $x(t)$ be the lowest at $t=T$, i.e., $t_{\frac{1}{x}}=T$. Depending on signs of $\alpha$ and $1-\beta\tau$ in Eq.~\eqref{eq:varying_g_y}, we consider the following four cases:
\begin{enumerate}
    \item If $\alpha>0$ and $\beta\tau<1$, $y(t)$ in Eq.~\eqref{eq:varying_g_y} is described essentially in the same way as Eq.~\eqref{eq:const_g_y}, while extra constants in Eq.~\eqref{eq:varying_g_y} do not affect the way to determine a phase difference between $x(t)$ and $y(t)$. Therefore, $t_y$ still follows
        \begin{equation}
            t_{-\frac{x'}{x}} \leq t_y\leq t_{\frac{1}{x}}=T,
        \end{equation}
        and the phase difference between $x(t)$ and $y(t)$ is determined in a similar way to the case with constant $g_{\rm A}(t)$ (i.e., $g_{\rm A}(t)=g$). $y(t)$ in this case will be called $y_1(t)$.
    \item If $\alpha>0$ and $\beta\tau>1$, $y(t)$ is determined in a similar way to $y_1(t)$, but with the flipped sign of $x'(t)$. Therefore,
        \begin{equation}
            0\leq t_y\leq t_{\frac{x'}{x}}.
        \end{equation}
        $y(t)$ in this case will be called $y_2(t)$.
    \item If $\alpha<0$ and $\beta\tau<1$, $y(t)$ is described in a similar way to $-y_2(t)$. Therefore,
        \begin{equation}
            t_x\leq t_y\leq t_{-\frac{x'}{x}}.
        \end{equation}
    \item If $\alpha<0$ and $\beta\tau>1$, $y(t)$ is described in a similar way to $-y_1(t)$. Therefore,
        \begin{equation}
            t_{\frac{x'}{x}}\leq t_y\leq t_x.
        \end{equation}
\end{enumerate}
In addition, both $x(t)$ and $y(t)$ can have symmetric waveforms as long as $|\alpha/(1-\beta\tau)| \gg \max_t |x'(t)|$ (for example, this condition can be satisfied when $\beta\tau\approx 1$).

Besides the case of Eq.~\eqref{eq:mammal_g_assume} with $\tau\ll T$, we analyze the case with $\tau\sim T/2$. In this case, $\tau=T/2+\epsilon$ with $|\epsilon|\ll T$, and $x(t+\tau)$ in Eq.~\eqref{eq:mammal_g_assume} can be approximated as
\begin{equation}
    x\left(t+\frac{T}{2}+\epsilon \right) \approx x\left(t+\frac{T}{2}\right)+\epsilon x'\left(t+ \frac{T}{2}\right).
\end{equation}
We further assume the waveform that $x(t+T/2) \approx J-x(t)$, where $J$ is a constant satisfying $J \approx (2/T)\int_0^T x(t)dt$. From Eq.~\eqref{eq:eq_animal_y},
\begin{equation}
    ky(t) \approx \frac{\alpha+\beta J - (1+\beta\epsilon) x'(t)}{x(t)} -(r_0+\beta).
\end{equation}
This equation takes a similar form to Eq.~\eqref{eq:varying_g_y}. By dividing four different categories of $y(t)$ depending on signs of $\alpha+\beta J$ and $1+\beta\epsilon$, it is straightforward to obtain similar results to our previous analysis of a phase difference between $x(t)$ and $y(t)$ when $\tau\ll T$.

In addition, both $x(t)$ and $y(t)$ can have symmetric waveforms as long as $|(\alpha+\beta J)/(1+\beta\epsilon)| \gg \max_t |x'(t)|$ (for example, this condition can be satisfied when $\beta\epsilon\approx -1$).

To illustrate the diverse phase differences conferred by BMAL1 cycling, we study the case with a sinusoidal wave $x(t)$ in Eq.~\eqref{eq:sine_wave} and consider the oscillation of $g_{\rm A}(t)$ in Eq.~\eqref{eq:mammal_g_assume}. From Eq.~\eqref{eq:eq_animal_y}, $t_y$ is obtained as
\begin{eqnarray}
    \omega t_y & = &2\pi n +2\tan^{-1}\left\{
        \frac{ \omega\alpha +C\beta[1-\cos(\omega\tau)] } { (C+L)[\omega-\beta\sin(\omega\tau)] } \right. \nonumber \\
     && - \left. \frac{ \sqrt{(\omega\alpha +C\beta[1-\cos(\omega\tau)])^2+(C^2-L^2)[\omega-\beta\sin(\omega\tau)]^2} } { (C+L)[\omega-\beta\sin(\omega\tau)] } \right\}
    \label{eq:mammal_sine_ty}
\end{eqnarray}
with $C=h_0\omega+L$ and an integer $n$ that satisfies $0<\omega t_y\leq 2\pi$. From Eq.~\eqref{eq:mammal_alpha_min}, $\alpha$ satisfies
\begin{equation}
    \alpha \geq \alpha_{\rm min}=
    \max\left\{L\sqrt{1-\frac{2\beta}{\omega} \sin(\omega\tau) +\frac{\beta^2}{\omega^2}} -\frac{\beta L}{\omega}, 0 \right\} -\beta h_0.
    \label{eq:mammal_sine_alpha_min}
\end{equation}
Equation~\eqref{eq:mammal_sine_ty} and $t_x=T/2$ give rise to the exact solution of the phase difference $\phi$ between $x(t)$ and $y(t)$ ($\phi=|\omega(t_x-t_y)|$), as plotted in Fig.~5e. This exact solution is in good agreement with our generic results based on the approximation Eqs.~\eqref{eq:x_t_tau_expansion} and~\eqref{eq:varying_g_y}.

\subsection{Data availability.} 

All relevant data are available in Methods, Figs.~2--4, Supplementary Methods, Supplementary Figs.~1--5, and Supplementary Data 1.

\subsection{Code availability.} 

Source codes for analyzing data in the manuscript have been deposited into the public repositories GitHub (https://github.com/h2jo/Circadian\_Waveform\_Analysis) and Zenodo (https://dx.doi.org/10.5281/zenodo.1466157).

\end{methods}

\newpage

\begin{addendum}
 \item This work was supported by Basic Science Research Program through the National Research Foundation of Korea (NRF) Grant 2015R1D1A1A01058958 funded by the Ministry of Education (H.-H.J.) and by the National Research Foundation of Korea (NRF) Grants N01160447 funded by the Ministry of Science, ICT and Future Planning (J.K.K.). This work was also supported by National Institutes of Health grant R01GM093285 and by a grant from the Next-Generation BioGreen21 Program (Systems and Synthetic Agrobiotech Center, Project PJ01327305), Rural Development Administration, Republic of Korea (D.E.S). P.-J.K. acknowledges the support by Hong Kong Baptist University, Research Committee, Start-up Grant for New Academics.
 \item[Author Contributions:] H.-H.J., Y.J.K., J.K.K., D.E.S., and P.-J.K. designed the research. H.-H.J., Y.J.K., M.F., and P.-J.K. performed the research. H.-H.J., Y.J.K., J.K.K., D.E.S., and P.-J.K. analyzed the data. H.-H.J., Y.J.K., J.K.K., D.E.S., and P.-J.K. wrote the manuscript.
 \item[Competing Interests:] The authors declare no competing financial or non-financial interests.
\end{addendum}

\newpage

\section*{Figures and Table}

\begin{figure}
    \begin{center}
    \includegraphics[width=0.66\columnwidth]{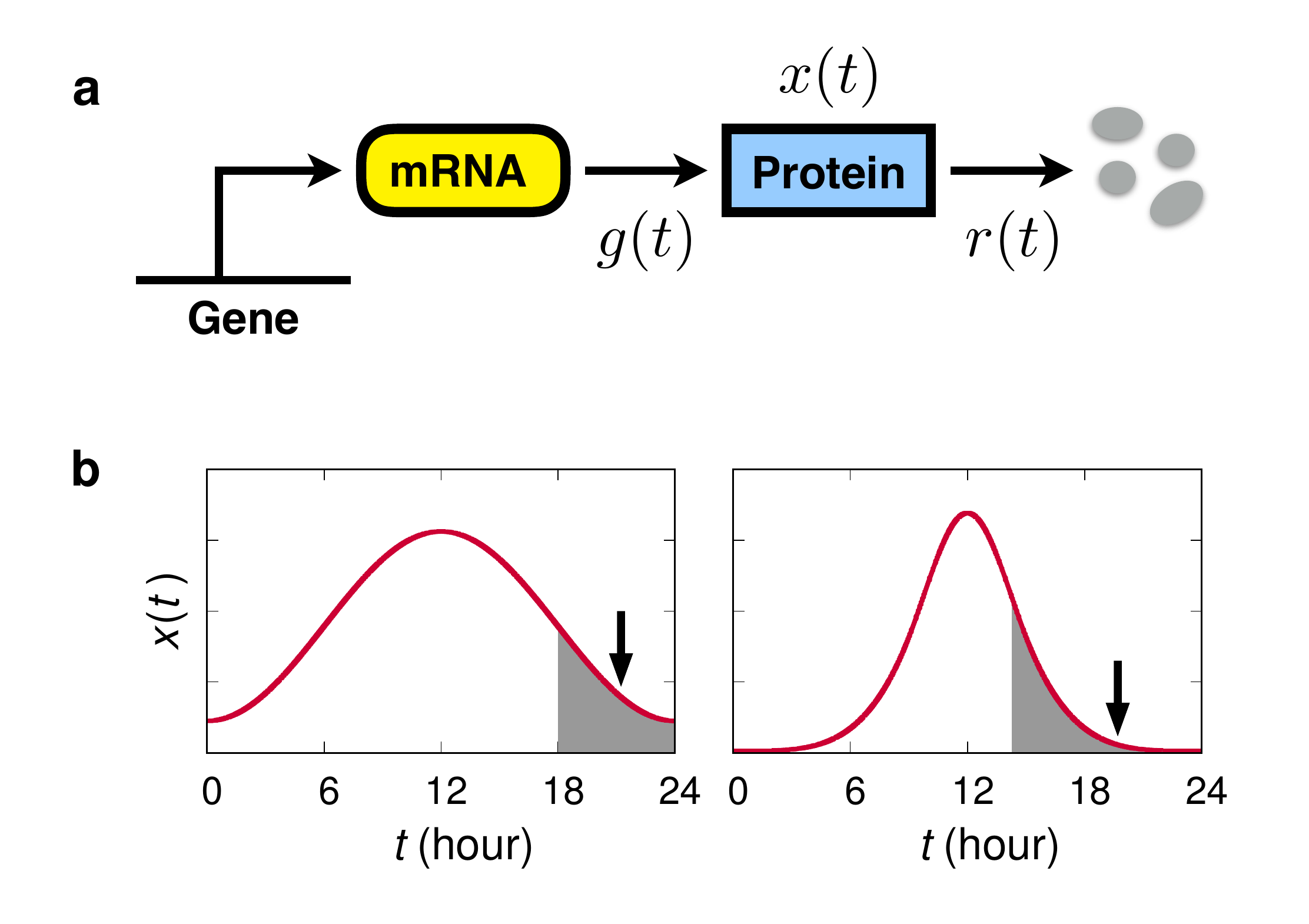}
    \end{center}
    \caption{Schematic diagrams of protein synthesis and turnover, and the resulting protein profiles in the circadian system. (\textbf{a}) Proteins are synthesized through mRNA-to-protein translation, and destined for degradation. (\textbf{b}) Cyclic protein abundances are represented by waveforms. For each waveform, the arrow indicates the point when $R(t)=r_{\rm min}$ in Eq.~\eqref{eq:rmin}, and the shaded area corresponds to the interval between the steepest decline and the trough. The right waveform ($r_{\rm min}\approx 0.69$~h$^{-1}$) has larger $r_{\rm min}$ than the left waveform ($r_{\rm min}\approx 0.30$~h$^{-1}$). For the definition of each notation in \textbf{a} and \textbf{b}, refer to Eq.~\eqref{eq:eq_plant}.}
\end{figure}

\newpage
\begin{figure}
    \begin{center}
    \includegraphics[width=0.66\columnwidth]{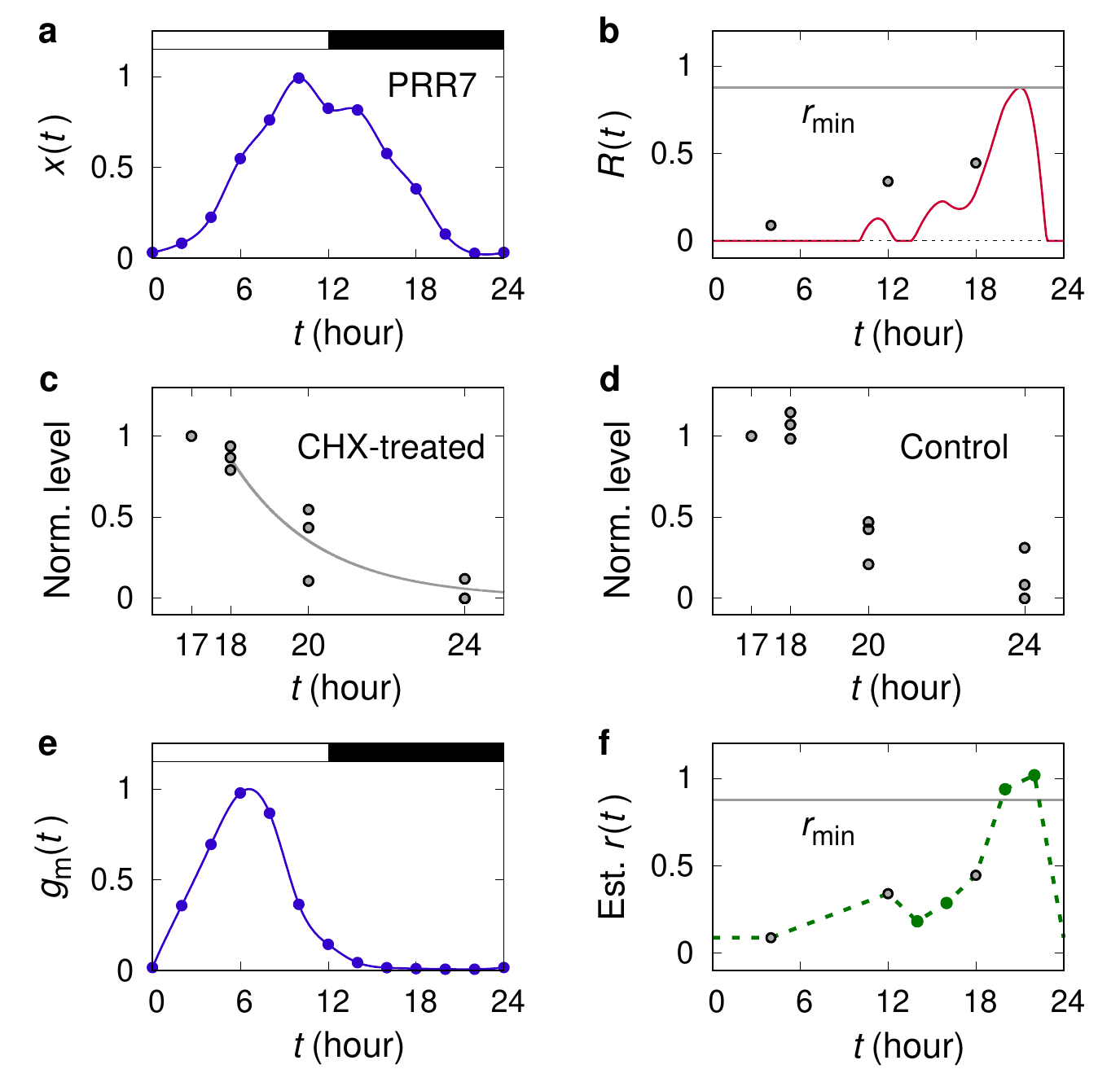}
    \end{center}
    \caption{PRR7 in the plant clock. For the definition of each notation, refer to Eqs.~\eqref{eq:eq_plant}, \eqref{eq:rt_condition}, \eqref{eq:rmin}, and~\eqref{eq:eq_plant_mRNA}. (\textbf{a}) Existing experimental data of PRR7 protein levels ($x(t)$, filled circles; normalized by the peak level of their spline curve)\cite{Nakamichi2010PseudoResponse}. (\textbf{b}) $R(t)$ (red solid line; calculated from $x(t)$ in \textbf{a}), $r_{\rm min}$ (gray solid line), and experimental $r(t)$ values (circles). The vertical axis unit is h$^{-1}$. The value of $r(t)$ at $t=18$~h is from our own experimental data in \textbf{c}. The rest $r(t)$ values in \textbf{b} are from previous experimental data\cite{Farre2007PRR7}. In agreement with Eq.~\eqref{eq:rt_condition}, there exists no $r(t)$ smaller than $R(t)$. (\textbf{c}) Our experimental measurement of PRR7 levels after CHX treatment at $t=17$~h. (\textbf{d}) Similar to \textbf{c}, but without CHX treatment. In \textbf{c} and \textbf{d}, PRR7 levels are normalized to the levels at $t=17$~h. Data points were obtained from three biological repeats. In \textbf{c}, considering a lag time for the full effect of CHX, an exponential fit (gray solid line) was made from $t=18$~h, and then $r(t) \approx 0.45 \pm 0.11$~h$^{-1}$ (avg. $\pm$ s.d.) at $t=18$~h in \textbf{b} was obtained (this standard deviation of $r(t)$ does not much change the cost reduction in Table 1, because it leads to $(c_{\rm g}-c)/c_{\rm g}\approx 0.68\sim 0.73$); an exponential fit from $t=17$~h also supports Eq.~\eqref{eq:rt_condition} (Supplementary Fig.~2). Control PRR7 levels at and after $t=18$~h in \textbf{d}, when averaged over three repeats at each time point and then rescaled together, are almost identical to $x(t)$ in \textbf{a}. (\textbf{e}) Existing experimental data of \emph{PRR7} mRNA levels ($g_{\rm m}(t)$, filled circles; normalized by the peak level of their spline curve)\cite{Flis2015Defining}. (\textbf{f}) Estimated $r(t)$ over time (green dashed line; green circles for direct calculation from experimental $r(t)$, $x(t)$, and $g_{\rm m}(t)$ using Eqs.~\eqref{eq:eq_plant_mRNA}--\eqref{eq:eq_plant_mRNA_rt} with constant $k$), along with $r_{\rm min}$ in \textbf{b}. The vertical axis unit is h$^{-1}$. All experimental data here pertain to 12L:12D cycles, and white and black segments in \textbf{a} and \textbf{e} correspond to light and dark intervals, respectively.}
\end{figure}

\newpage
\begin{figure}
    \begin{center}
    \includegraphics[width=0.66\columnwidth]{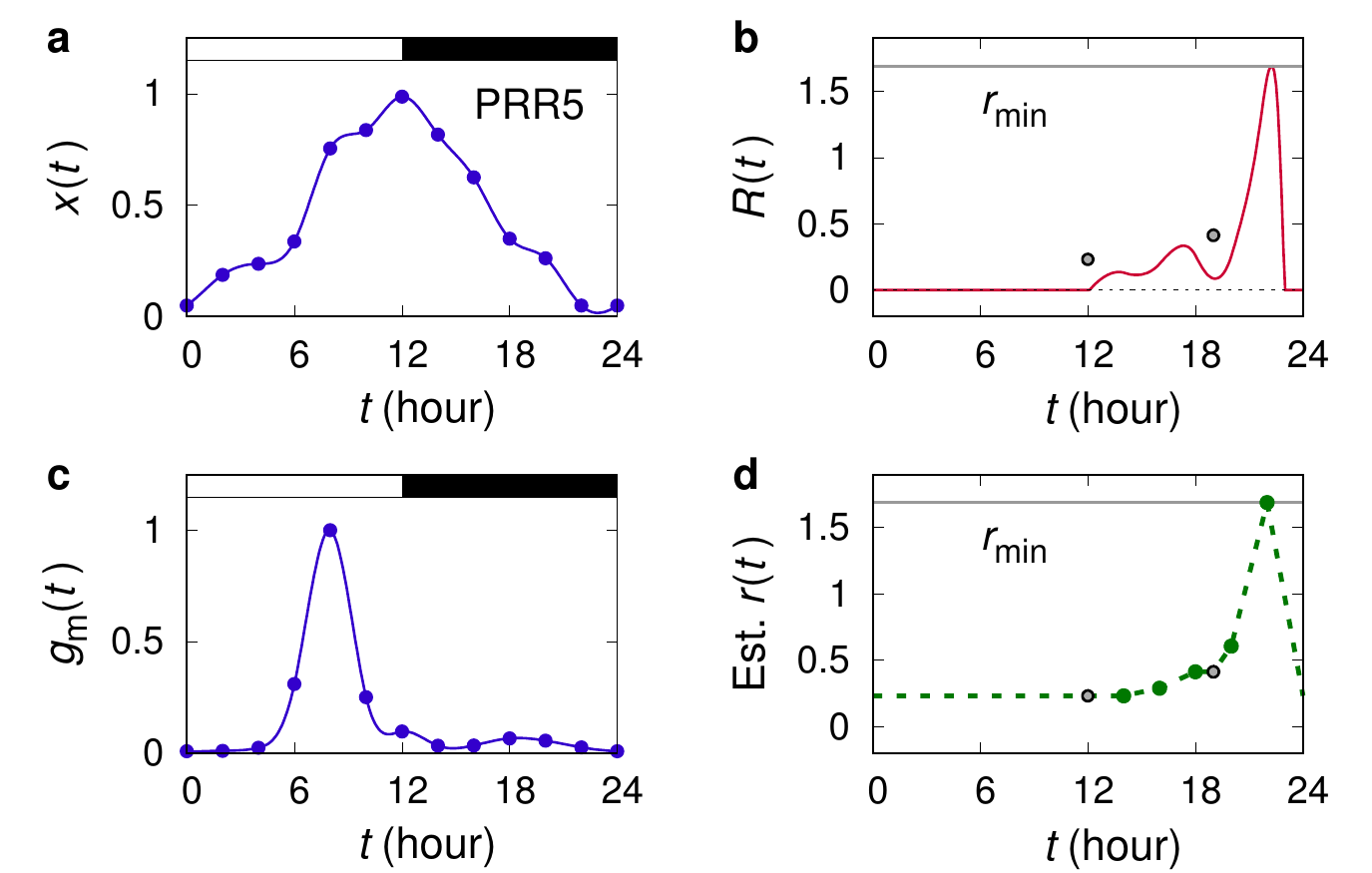}
    \end{center}
    \caption{PRR5 in the plant clock. For the definition of each notation, refer to Eqs.~\eqref{eq:eq_plant}, \eqref{eq:rt_condition}, \eqref{eq:rmin}, and~\eqref{eq:eq_plant_mRNA}. (\textbf{a}) Existing experimental data of PRR5 protein levels ($x(t)$, filled circles; normalized by the peak level of their spline curve)\cite{Nakamichi2010PseudoResponse}. (\textbf{b}) $R(t)$ (red solid line; calculated from $x(t)$ in \textbf{a}), $r_{\rm min}$ (gray solid line), and empirical $r(t)$ values (circles). The vertical axis unit is h$^{-1}$. The $r(t)$ values are from previous experimental data\cite{Baudry2010FBox, Wang2010PRR5}. In agreement with Eq.~\eqref{eq:rt_condition}, there exists no $r(t)$ smaller than $R(t)$. (\textbf{c}) Existing experimental data of \emph{PRR5} mRNA levels ($g_{\rm m}(t)$, filled circles; normalized by the peak level of their spline curve)\cite{Flis2015Defining}. (\textbf{d}) Estimated $r(t)$ over time (green dashed line; green circles for direct calculation from experimental $r(t)$, $x(t)$, and $g_{\rm m}(t)$ using Eqs.~\eqref{eq:eq_plant_mRNA}--\eqref{eq:eq_plant_mRNA_rt} with constant $k$), along with $r_{\rm min}$ in \textbf{b}. The vertical axis unit is h$^{-1}$. All experimental data here pertain to 12L:12D cycles, except for $r(t)$ at $t=19$~h in \textbf{b}, which was collected from a different light condition due to the scarcity of experimental data (Supplementary Methods). White and black segments in \textbf{a} and \textbf{c} correspond to light and dark intervals, respectively.}
\end{figure}

\newpage
\begin{figure}
    \begin{center}
    \includegraphics[width=\textwidth]{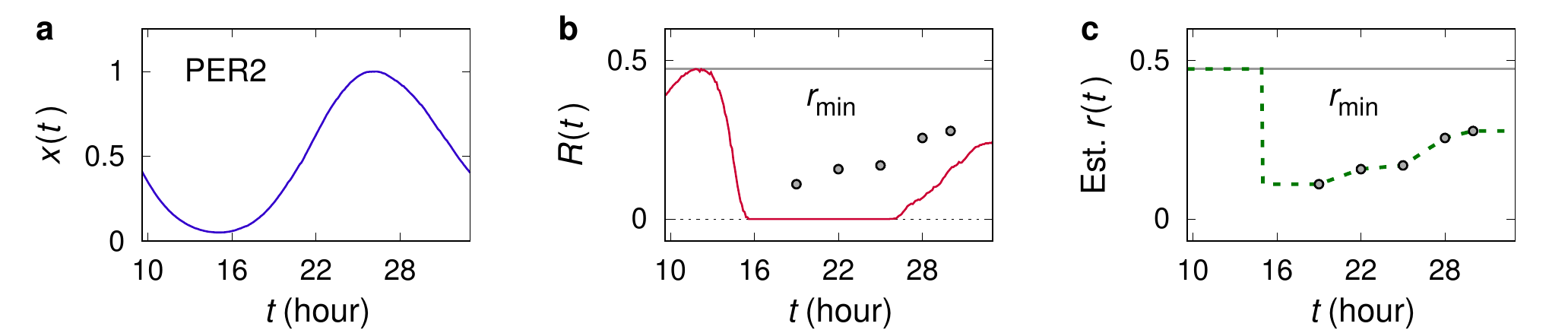}
    \end{center}
    \caption{PER2 in the mammalian clock. For the definition of each notation, refer to Eqs.~\eqref{eq:eq_plant}, \eqref{eq:rt_condition}, and~\eqref{eq:rmin}. (\textbf{a}) Existing experimental data of PER2 protein levels ($x(t)$, normalized by the peak level; moving window average of experimental data)\cite{Zhou2015Period2}. (\textbf{b}) $R(t)$ (red solid line; calculated from $x(t)$ in \textbf{a}), $r_{\rm min}$ (gray solid line), and empirical $r(t)$ values (circles). The vertical axis unit is h$^{-1}$. The $r(t)$ values are from previous experimental data\cite{Zhou2015Period2}. In agreement with Eq.~\eqref{eq:rt_condition}, there exists no $r(t)$ smaller than $R(t)$. (\textbf{c}) Estimated $r(t)$ over time (green dashed line; green circles for experimental $r(t)$ data in \textbf{b}), along with $r_{\rm min}$ in \textbf{b}. The vertical axis unit is h$^{-1}$.}
\end{figure}

\newpage
\begin{figure}
    \begin{center}
    \includegraphics[width=0.66\columnwidth]{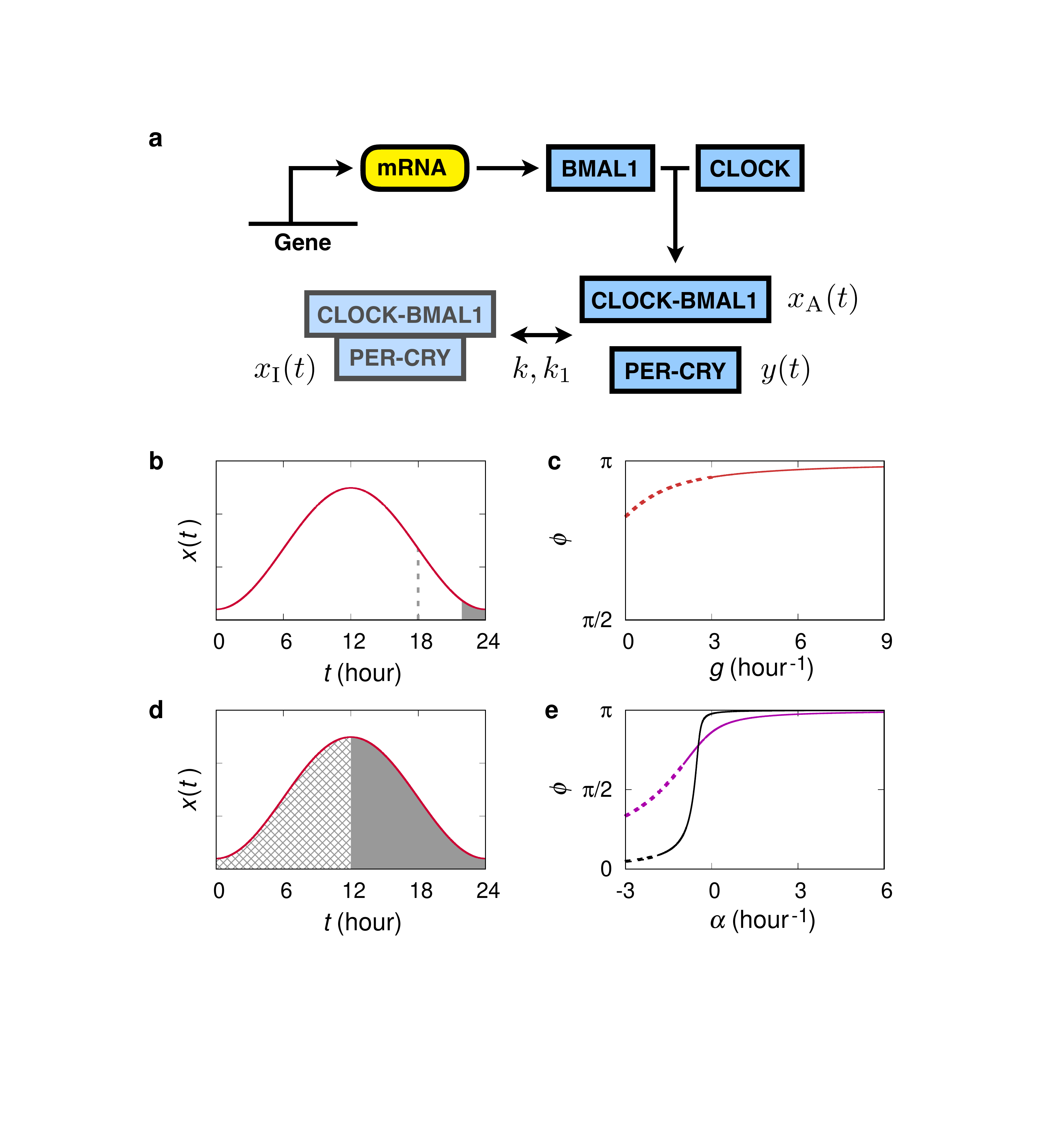}
    \end{center}
    \caption{
        Protein activity dynamics of the mammalian circadian system. (\textbf{a}) CLOCK and BMAL1 proteins, as well as PER and CRY proteins, heterodimerize with each other. The activities of CLOCK-BMAL1 complexes are inhibited by PER-CRY complexes through their interactions in the nucleus. For the definition of each notation, refer to Eq.~\eqref{eq:eq_animal} and its predecessor equations. (\textbf{b}--\textbf{e}) Possible phase relationship between active CLOCK-BMAL1, and PER-CRY that is not binding to CLOCK-BMAL1. (\textbf{b}) Regarding Eq.~\eqref{eq:eq_animal_y}, when $g_{\rm A}(t)$ is constant, the shaded area corresponds to a range of $y(t)$'s peak time, i.e., $t_y$'s range in Eq.~\eqref{eq:t_y_t_R}. For comparison, the dashed line indicates the time of $x(t)$'s steepest decline. (\textbf{c}) Phase difference $\phi$ between $x(t)$ and $y(t)$ as a function of $g$, when $g_{\rm A}(t)=g$ and $x(t)$ is modeled by Eq.~\eqref{eq:sine_wave} with $L=3$~h$^{-1}$ and $h_0=2$. Dotted is an infeasible solution with $g<g_{\rm min}=L$ (Eq.~\eqref{eq:g_min}). (\textbf{d}) Regarding Eq.~\eqref{eq:eq_animal_y}, when $g_{\rm A}(t)$ varies over time as in Eq.~\eqref{eq:mammal_g_assume}, the left and right shaded areas correspond to the ranges of $y(t)$'s peak time for $\beta\tau>1$ and for $\beta\tau<1$, respectively (Methods). (\textbf{e}) Phase difference $\phi$ between $x(t)$ and $y(t)$ as a function of $\alpha$, when $g_{\rm A}(t)$ varies over time as in Eq.~\eqref{eq:mammal_g_assume} with $\beta=0.5$~h$^{-1}$ (violet) or $\beta=0.95$~h$^{-1}$ (black), and $\tau=1$~h, and $x(t)$ is modeled by Eq.~\eqref{eq:sine_wave} with $L=3$~h$^{-1}$ and $h_0=2$. Dotted is an infeasible solution with $\alpha<\alpha_{\rm min}$. Full details are described in Methods.}
\end{figure}

\newpage
\begin{table}
    \begin{center}
    \begin{tabular}{lcccc}
        \hline
        Protein & $r_{\rm min}$ (h$^{-1}$) & $c_{\rm g}$ (h$^{-1}$) & $c$ (h$^{-1}$) & Cost reduction \\ \hline
        PRR7 & 0.88 & 0.40 & 0.12 & $\sim$70\% \\ 
        PRR5 & 1.69 & 0.77 & 0.13 & $\sim$83\% \\ 
        PER2 & 0.47 & 0.23 & 0.11 & $\sim$52\% \\ \hline
    \end{tabular}
    \end{center}
    \caption{Estimated values of $r_{\rm min}$, $c_{\rm g} = r_{\rm min}\langle x(t)\rangle$, and $c=\langle r(t)x(t)\rangle$ as well as cost reduction for PRR7, PRR5, and PER2. For the definitions of $c_{\rm g}$ and $c$, refer to Eqs.~\eqref{eq:cost_define} and~\eqref{eq:global_cost_define}. The cost reduction due to the time- or phase-specific $r(t)$ is defined as $(c_{\rm g}-c)/c_{\rm g}$. We here assume constant $k$ in Eq.~\eqref{eq:eq_plant_mRNA}. We treat $x(t)$ as dimensionless through the normalization of $x(t)$ by its peak value (Figs.~2a, 3a, and 4a), and thus units of $r_{\rm min}$, $c_{\rm g}$, and $c$ in the Table are hour$^{-1}$. The cost reduction itself is not a quantity affected by the normalization of $x(t)$, and hence there is no loss of generality in its values.}
\end{table}


\section*{References}
\begin{thebibliography}{10}
\expandafter\ifx\csname url\endcsname\relax
  \def\url#1{\texttt{#1}}\fi
\expandafter\ifx\csname urlprefix\endcsname\relax\def\urlprefix{URL }\fi
\providecommand{\bibinfo}[2]{#2}
\providecommand{\eprint}[2][]{\url{#2}}

\bibitem{Nagel2012Complexity}
\bibinfo{author}{Nagel, D.~H.} \& \bibinfo{author}{Kay, S.~A.} \newblock \bibinfo{title}{Complexity in the wiring and regulation of plant circadian networks}.  \newblock \emph{\bibinfo{journal}{Current Biology}} \textbf{\bibinfo{volume}{22}}, \bibinfo{pages}{R648--R657} (\bibinfo{year}{2012}).  \newblock \urlprefix\url{http://dx.doi.org/10.1016/j.cub.2012.07.025}.

\bibitem{Sehgal1994Loss}
\bibinfo{author}{Sehgal, A.}, \bibinfo{author}{Price, J.~L.},
  \bibinfo{author}{Man, B.} \& \bibinfo{author}{Young, M.~W.}
  \newblock \bibinfo{title}{Loss of circadian behavioral rhythms and \emph{per} RNA oscillations in the \emph{Drosophila} mutant \emph{timeless}}.
\newblock \emph{\bibinfo{journal}{Science}} \textbf{\bibinfo{volume}{263}},
  \bibinfo{pages}{1603--1606} (\bibinfo{year}{1994}).
\newblock \urlprefix\url{http://dx.doi.org/10.1126/science.8128246}.

\bibitem{Brody1973Circadian}
\bibinfo{author}{Brody, S.} \& \bibinfo{author}{Harris, S.}
\newblock \bibinfo{title}{Circadian rhythms in \emph{Neurospora}: Spatial differences in pyridine nucleotide levels}.
\newblock \emph{\bibinfo{journal}{Science}} \textbf{\bibinfo{volume}{180}},
  \bibinfo{pages}{498--500} (\bibinfo{year}{1973}).
\newblock \urlprefix\url{http://dx.doi.org/10.1126/science.180.4085.498}.

\bibitem{Gachon2004Mammalian}
\bibinfo{author}{Gachon, F.}, \bibinfo{author}{Nagoshi, E.},
  \bibinfo{author}{Brown, S.}, \bibinfo{author}{Ripperger, J.} \&
  \bibinfo{author}{Schibler, U.}
\newblock \bibinfo{title}{The mammalian circadian timing system: from gene
  expression to physiology}.
\newblock \emph{\bibinfo{journal}{Chromosoma}} \textbf{\bibinfo{volume}{113}},
  \bibinfo{pages}{103--112} (\bibinfo{year}{2004}).
\newblock \urlprefix\url{http://dx.doi.org/10.1007/s00412-004-0296-2}.

\bibitem{Fei2018Design}
\bibinfo{author}{Fei, C.}, \bibinfo{author}{Cao, Y.}, \bibinfo{author}{Ouyang, Q.} \& \bibinfo{author}{Tu, Y.} \newblock \bibinfo{title}{Design principles for enhancing phase sensitivity and suppressing phase fluctuations simultaneously in biochemical oscillatory systems}.  \newblock \emph{\bibinfo{journal}{Nature Communications}} \textbf{\bibinfo{volume}{9}}, \bibinfo{pages}{1434} (\bibinfo{year}{2018}).  \newblock \urlprefix\url{http://dx.doi.org/10.1038/s41467-018-03826-4}.

\bibitem{Hsu2014Wheels}
\bibinfo{author}{Hsu, P.~Y.} \& \bibinfo{author}{Harmer, S.~L.}
\newblock \bibinfo{title}{Wheels within wheels: the plant circadian system}.
\newblock \emph{\bibinfo{journal}{Trends in Plant Science}}
  \textbf{\bibinfo{volume}{19}}, \bibinfo{pages}{240--249}
  (\bibinfo{year}{2014}).
\newblock \urlprefix\url{http://dx.doi.org/10.1016/j.tplants.2013.11.007}.

\bibitem{Hurley2015Dissecting}
\bibinfo{author}{Hurley, J.}, \bibinfo{author}{Loros, J.~J.} \&
  \bibinfo{author}{Dunlap, J.~C.}
\newblock \bibinfo{title}{Dissecting the mechanisms of the clock in
  \emph{Neurospora}}.
\newblock \emph{\bibinfo{journal}{Methods in Enzymology}},
  vol. \bibinfo{volume}{551}, \bibinfo{pages}{29--52}
  (\bibinfo{publisher}{Elsevier}, \bibinfo{year}{2015}).
\newblock \urlprefix\url{http://dx.doi.org/10.1016/bs.mie.2014.10.009}.

\bibitem{MendozaViveros2017Molecular}
\bibinfo{author}{Mendoza-Viveros, L.} \emph{et~al.}
\newblock \bibinfo{title}{Molecular modulators of the circadian clock: lessons
  from flies and mice}.
\newblock \emph{\bibinfo{journal}{Cellular and Molecular Life Sciences}}
  \textbf{\bibinfo{volume}{74}}, \bibinfo{pages}{1035--1059}
  (\bibinfo{year}{2017}).
\newblock \urlprefix\url{http://dx.doi.org/10.1007/s00018-016-2378-8}.

\bibitem{Sancar2015Dawn}
\bibinfo{author}{Sancar, C.}, \bibinfo{author}{Sancar, G.},
  \bibinfo{author}{Ha, N.}, \bibinfo{author}{Cesbron, F.} \&
  \bibinfo{author}{Brunner, M.}
\newblock \bibinfo{title}{Dawn- and dusk-phased circadian transcription rhythms
  coordinate anabolic and catabolic functions in \emph{Neurospora}}.
\newblock \emph{\bibinfo{journal}{BMC Biology}} \textbf{\bibinfo{volume}{13}},
  \bibinfo{pages}{17} (\bibinfo{year}{2015}).
\newblock \urlprefix\url{http://dx.doi.org/10.1186/s12915-015-0126-4}.

\bibitem{Lee2001Posttranslational}
\bibinfo{author}{Lee, C.}, \bibinfo{author}{Etchegaray, J.~P.},
  \bibinfo{author}{Cagampang, F.~R.}, \bibinfo{author}{Loudon, A.~S.} \&
  \bibinfo{author}{Reppert, S.~M.}
\newblock \bibinfo{title}{Posttranslational mechanisms regulate the mammalian
  circadian clock.}
\newblock \emph{\bibinfo{journal}{Cell}} \textbf{\bibinfo{volume}{107}},
  \bibinfo{pages}{855--867} (\bibinfo{year}{2001}).
\newblock \urlprefix\url{http://dx.doi.org/10.1016/S0092-8674(01)00610-9}.

\bibitem{Liu2008Redundant}
\bibinfo{author}{Liu, A.~C.} \emph{et~al.}
\newblock \bibinfo{title}{Redundant function of REV-ERB$\alpha$ and $\beta$ and non-essential role for Bmal1 cycling in transcriptional regulation of
  intracellular circadian rhythms}.
\newblock \emph{\bibinfo{journal}{PLOS Genetics}} \textbf{\bibinfo{volume}{4}},
  \bibinfo{pages}{e1000023} (\bibinfo{year}{2008}).
\newblock \urlprefix\url{http://dx.doi.org/10.1371/journal.pgen.1000023}.

\bibitem{Preitner2002Orphan}
\bibinfo{author}{Preitner, N.} \emph{et~al.}
\newblock \bibinfo{title}{The orphan nuclear receptor REV-ERB$\alpha$ controls
  circadian transcription within the positive limb of the mammalian circadian
  oscillator}.
\newblock \emph{\bibinfo{journal}{Cell}} \textbf{\bibinfo{volume}{110}},
  \bibinfo{pages}{251--260} (\bibinfo{year}{2002}).
\newblock \urlprefix\url{http://dx.doi.org/10.1016/s0092-8674(02)00825-5}.

\bibitem{McDearmon2006Dissecting}
\bibinfo{author}{McDearmon, E.~L.} \emph{et~al.}
\newblock \bibinfo{title}{Dissecting the functions of the mammalian clock
  protein BMAL1 by tissue-specific rescue in mice}.
\newblock \emph{\bibinfo{journal}{Science}} \textbf{\bibinfo{volume}{314}},
  \bibinfo{pages}{1304--1308} (\bibinfo{year}{2006}).
\newblock \urlprefix\url{http://dx.doi.org/10.1126/science.1132430}.

\bibitem{Kiba2007Targeted}
\bibinfo{author}{Kiba, T.}, \bibinfo{author}{Henriques, R.},
  \bibinfo{author}{Sakakibara, H.} \& \bibinfo{author}{Chua, N.-H.}
\newblock \bibinfo{title}{Targeted degradation of PSEUDO-RESPONSE REGULATOR5 by an SCF$^{\rm ZTL}$ complex regulates clock function and photomorphogenesis in \emph{Arabidopsis thaliana}}.
\newblock \emph{\bibinfo{journal}{The Plant Cell}}
  \textbf{\bibinfo{volume}{19}}, \bibinfo{pages}{2516--2530}
  (\bibinfo{year}{2007}).
\newblock \urlprefix\url{http://dx.doi.org/10.1105/tpc.107.053033}.

\bibitem{Fujiwara2008Posttranslational}
\bibinfo{author}{Fujiwara, S.} \emph{et~al.}
\newblock \bibinfo{title}{Post-translational regulation of the \emph{Arabidopsis} circadian clock through selective proteolysis and phosphorylation of pseudo-response regulator proteins}.
\newblock \emph{\bibinfo{journal}{Journal of Biological Chemistry}}
  \textbf{\bibinfo{volume}{283}}, \bibinfo{pages}{23073--23083}
  (\bibinfo{year}{2008}).
\newblock \urlprefix\url{http://dx.doi.org/10.1074/jbc.m803471200}.

\bibitem{Foo2016Kernel}
\bibinfo{author}{Foo, M.}, \bibinfo{author}{Somers, D.~E.} \&
  \bibinfo{author}{Kim, P.-J.}
\newblock \bibinfo{title}{Kernel architecture of the genetic circuitry of the \emph{Arabidopsis} circadian system}.
\newblock \emph{\bibinfo{journal}{PLOS Computational Biology}}
  \textbf{\bibinfo{volume}{12}}, \bibinfo{pages}{e1004748}
  (\bibinfo{year}{2016}).
\newblock \urlprefix\url{http://dx.doi.org/10.1371/journal.pcbi.1004748}.

\bibitem{Hatakeyama2015Reciprocity}
\bibinfo{author}{Hatakeyama, T.~S.} \& \bibinfo{author}{Kaneko, K.}
\newblock \bibinfo{title}{Reciprocity between robustness of period and plasticity of phase in biological clocks}.
\newblock \emph{\bibinfo{journal}{Physical Review Letters}}
  \textbf{\bibinfo{volume}{115}}, \bibinfo{pages}{218101}
  (\bibinfo{year}{2015}).
\newblock \urlprefix\url{http://dx.doi.org/10.1103/physrevlett.115.218101}.

\bibitem{Luck2014Rhythmic}
\bibinfo{author}{L\"{u}ck, S.}, \bibinfo{author}{Thurley, K.},
  \bibinfo{author}{Thaben, P.~F.} \& \bibinfo{author}{Westermark, P.~O.}
\newblock \bibinfo{title}{Rhythmic degradation explains and unifies circadian transcriptome and proteome data}.
\newblock \emph{\bibinfo{journal}{Cell Reports}} \textbf{\bibinfo{volume}{9}},
  \bibinfo{pages}{741--751} (\bibinfo{year}{2014}).
\newblock \urlprefix\url{http://dx.doi.org/10.1016/j.celrep.2014.09.021}.

\bibitem{Patton2016Combined}
\bibinfo{author}{Patton, A.~P.}, \bibinfo{author}{Chesham, J.~E.} \&
  \bibinfo{author}{Hastings, M.~H.}
\newblock \bibinfo{title}{Combined pharmacological and genetic manipulations unlock unprecedented temporal elasticity and reveal phase-specific modulation of the molecular circadian clock of the mouse suprachiasmatic nucleus}.
\newblock \emph{\bibinfo{journal}{The Journal of Neuroscience}}
  \textbf{\bibinfo{volume}{36}}, \bibinfo{pages}{9326--9341}
  (\bibinfo{year}{2016}).
\newblock \urlprefix\url{http://dx.doi.org/10.1523/jneurosci.0958-16.2016}.

\bibitem{Korencic2014Timing}
\bibinfo{author}{Koren\v{c}i\v{c}, A.} \emph{et~al.}
\newblock \bibinfo{title}{Timing of circadian genes in mammalian tissues}.
\newblock \emph{\bibinfo{journal}{Scientific Reports}}
  \textbf{\bibinfo{volume}{4}}, \bibinfo{pages}{5782} (\bibinfo{year}{2014}).
\newblock \urlprefix\url{http://dx.doi.org/10.1038/srep05782}.

\bibitem{Yoo2013Competing}
\bibinfo{author}{Yoo, S.-H.} \emph{et~al.}
\newblock \bibinfo{title}{Competing E3 ubiquitin ligases govern circadian
  periodicity by degradation of CRY in nucleus and cytoplasm}.
\newblock \emph{\bibinfo{journal}{Cell}} \textbf{\bibinfo{volume}{152}},
  \bibinfo{pages}{1091--1105} (\bibinfo{year}{2013}).
\newblock \urlprefix\url{http://dx.doi.org/10.1016/j.cell.2013.01.055}.

\bibitem{Kim2012Existence}
\bibinfo{author}{Kim, J.~K.} \& \bibinfo{author}{Forger, D.~B.}
\newblock \bibinfo{title}{On the existence and uniqueness of biological clock models matching experimental data}.
\newblock \emph{\bibinfo{journal}{SIAM Journal on Applied Mathematics}}
  \textbf{\bibinfo{volume}{72}}, \bibinfo{pages}{1842--1855}
  (\bibinfo{year}{2012}).
\newblock \urlprefix\url{http://dx.doi.org/10.1137/120867809}.

\bibitem{deMontaigu2015Natural}
\bibinfo{author}{de~Montaigu, A.} \emph{et~al.}
\newblock \bibinfo{title}{Natural diversity in daily rhythms of gene expression contributes to phenotypic variation}.
\newblock \emph{\bibinfo{journal}{Proceedings of the National Academy of
  Sciences}} \textbf{\bibinfo{volume}{112}}, \bibinfo{pages}{905--910}
  (\bibinfo{year}{2015}).
\newblock \urlprefix\url{http://dx.doi.org/10.1073/pnas.1422242112}.

\bibitem{Scialdone2013Arabidopsis}
\bibinfo{author}{Scialdone, A.} \emph{et~al.}
\newblock \bibinfo{title}{\emph{Arabidopsis} plants perform arithmetic division to prevent starvation at night}.
\newblock \emph{\bibinfo{journal}{eLife}} \textbf{\bibinfo{volume}{2}},
  \bibinfo{pages}{e00669} (\bibinfo{year}{2013}).
\newblock \urlprefix\url{http://dx.doi.org/10.7554/elife.00669}.

\bibitem{Cole2017Brain}
\bibinfo{author}{Cole, S.~R.} \& \bibinfo{author}{Voytek, B.}
\newblock \bibinfo{title}{Brain oscillations and the importance of waveform shape}.
\newblock \emph{\bibinfo{journal}{Trends in Cognitive Sciences}}
  \textbf{\bibinfo{volume}{21}}, \bibinfo{pages}{137--149}
  (\bibinfo{year}{2017}).
\newblock \urlprefix\url{http://dx.doi.org/10.1016/j.tics.2016.12.008}.

\bibitem{Nakamichi2010PseudoResponse}
\bibinfo{author}{Nakamichi, N.} \emph{et~al.}
\newblock \bibinfo{title}{PSEUDO-RESPONSE REGULATORS 9, 7, and 5 are
  transcriptional repressors in the \emph{Arabidopsis} circadian clock}.
\newblock \emph{\bibinfo{journal}{The Plant Cell}}
  \textbf{\bibinfo{volume}{22}}, \bibinfo{pages}{594--605}
  (\bibinfo{year}{2010}).
\newblock \urlprefix\url{http://dx.doi.org/10.1105/tpc.109.072892}.

\bibitem{Flis2015Defining}
\bibinfo{author}{Flis, A.} \emph{et~al.}
\newblock \bibinfo{title}{Defining the robust behaviour of the plant clock gene circuit with absolute RNA timeseries and open infrastructure}.
\newblock \emph{\bibinfo{journal}{Open Biology}} \textbf{\bibinfo{volume}{5}},
  \bibinfo{pages}{150042} (\bibinfo{year}{2015}).
\newblock \urlprefix\url{http://dx.doi.org/10.1098/rsob.150042}.

\bibitem{Farre2007PRR7}
\bibinfo{author}{Farr\'{e}, E.~M.} \& \bibinfo{author}{Kay, S.~A.}
\newblock \bibinfo{title}{PRR7 protein levels are regulated by light and the
  circadian clock in \emph{Arabidopsis}}.
\newblock \emph{\bibinfo{journal}{The Plant Journal}}
  \textbf{\bibinfo{volume}{52}}, \bibinfo{pages}{548--560}
  (\bibinfo{year}{2007}).
\newblock \urlprefix\url{http://dx.doi.org/10.1111/j.1365-313x.2007.03258.x}.

\bibitem{Baudry2010FBox}
\bibinfo{author}{Baudry, A.} \emph{et~al.}
\newblock \bibinfo{title}{F-Box proteins FKF1 and LKP2 act in concert with
  ZEITLUPE to control \emph{Arabidopsis} clock progression}.
\newblock \emph{\bibinfo{journal}{The Plant Cell}}
  \textbf{\bibinfo{volume}{22}}, \bibinfo{pages}{606--622}
  (\bibinfo{year}{2010}).
\newblock \urlprefix\url{http://dx.doi.org/10.1105/tpc.109.072843}.

\bibitem{Wang2010PRR5}
\bibinfo{author}{Wang, L.}, \bibinfo{author}{Fujiwara, S.} \&
  \bibinfo{author}{Somers, D.~E.}
\newblock \bibinfo{title}{PRR5 regulates phosphorylation, nuclear import and subnuclear localization of TOC1 in the \emph{Arabidopsis} circadian clock}.
\newblock \emph{\bibinfo{journal}{The EMBO Journal}}
  \textbf{\bibinfo{volume}{29}}, \bibinfo{pages}{1903--1915}
  (\bibinfo{year}{2010}).
\newblock \urlprefix\url{http://dx.doi.org/10.1038/emboj.2010.76}.

\bibitem{Kim2003Circadian}
\bibinfo{author}{Kim, W.-Y.}, \bibinfo{author}{Geng, R.} \&
  \bibinfo{author}{Somers, D.~E.}
\newblock \bibinfo{title}{Circadian phase-specific degradation of the F-box protein ZTL is mediated by the proteasome}.
\newblock \emph{\bibinfo{journal}{Proceedings of the National Academy of
  Sciences}} \textbf{\bibinfo{volume}{100}}, \bibinfo{pages}{4933--4938}
  (\bibinfo{year}{2003}).
\newblock \urlprefix\url{http://dx.doi.org/10.1073/pnas.0736949100}.

\bibitem{Mas2003Targeted}
\bibinfo{author}{M\'{a}s, P.}, \bibinfo{author}{Kim, W.-Y.},
  \bibinfo{author}{Somers, D.~E.} \& \bibinfo{author}{Kay, S.~A.}
\newblock \bibinfo{title}{Targeted degradation of TOC1 by ZTL modulates
    circadian function in \emph{Arabidopsis thaliana}}.
\newblock \emph{\bibinfo{journal}{Nature}} \textbf{\bibinfo{volume}{426}},
  \bibinfo{pages}{567--570} (\bibinfo{year}{2003}).
\newblock \urlprefix\url{http://dx.doi.org/10.1038/nature02163}.

\bibitem{Schwanhausser2011Global}
\bibinfo{author}{Schwanh\"{a}usser, B.} \emph{et~al.}
\newblock \bibinfo{title}{Global quantification of mammalian gene expression control}.
\newblock \emph{\bibinfo{journal}{Nature}} \textbf{\bibinfo{volume}{473}},
  \bibinfo{pages}{337--342} (\bibinfo{year}{2011}).
\newblock \urlprefix\url{http://dx.doi.org/10.1038/nature10098}.

\bibitem{Christiano2014Global}
\bibinfo{author}{Christiano, R.}, \bibinfo{author}{Nagaraj, N.},
  \bibinfo{author}{Fr\"{o}hlich, F.} \& \bibinfo{author}{Walther, T.~C.}
  \newblock \bibinfo{title}{Global proteome turnover analyses of the yeasts \emph{S. cerevisiae} and \emph{S. pombe}}.
\newblock \emph{\bibinfo{journal}{Cell Reports}} \textbf{\bibinfo{volume}{9}},
  \bibinfo{pages}{1959--1965} (\bibinfo{year}{2014}).
\newblock \urlprefix\url{http://dx.doi.org/10.1016/j.celrep.2014.10.065}.

\bibitem{Zhou2015Period2}
\bibinfo{author}{Zhou, M.}, \bibinfo{author}{Kim, J.~K.}, \bibinfo{author}{Eng,
  G.~W.}, \bibinfo{author}{Forger, D.~B.} \& \bibinfo{author}{Virshup, D.~M.}
\newblock \bibinfo{title}{A Period2 phosphoswitch regulates and temperature
  compensates circadian period}.
\newblock \emph{\bibinfo{journal}{Molecular Cell}}
  \textbf{\bibinfo{volume}{60}}, \bibinfo{pages}{77--88}
  (\bibinfo{year}{2015}).
\newblock \urlprefix\url{http://dx.doi.org/10.1016/j.molcel.2015.08.022}.

\bibitem{Leloup2003Toward}
\bibinfo{author}{Leloup, J.-C.} \& \bibinfo{author}{Goldbeter, A.}
\newblock \bibinfo{title}{Toward a detailed computational model for the
  mammalian circadian clock}.
\newblock \emph{\bibinfo{journal}{Proceedings of the National Academy of
  Sciences}} \textbf{\bibinfo{volume}{100}}, \bibinfo{pages}{7051--7056}
  (\bibinfo{year}{2003}).
\newblock \urlprefix\url{http://dx.doi.org/10.1073/pnas.1132112100}.

\bibitem{Kim2012Mechanism}
\bibinfo{author}{Kim, J.~K.} \& \bibinfo{author}{Forger, D.~B.}
\newblock \bibinfo{title}{A mechanism for robust circadian timekeeping via
  stoichiometric balance}.
\newblock \emph{\bibinfo{journal}{Molecular Systems Biology}}
  \textbf{\bibinfo{volume}{8}}, \bibinfo{pages}{630} (\bibinfo{year}{2012}).
\newblock \urlprefix\url{http://dx.doi.org/10.1038/msb.2012.62}.

\bibitem{Vanselow2006Differential}
\bibinfo{author}{Vanselow, K.} \emph{et~al.}
\newblock \bibinfo{title}{Differential effects of PER2 phosphorylation:
  molecular basis for the human familial advanced sleep phase syndrome
  (FASPS)}.
\newblock \emph{\bibinfo{journal}{Genes \& Development}}
  \textbf{\bibinfo{volume}{20}}, \bibinfo{pages}{2660--2672}
  (\bibinfo{year}{2006}).
\newblock \urlprefix\url{http://dx.doi.org/10.1101/gad.397006}.

\bibitem{vanOoijen2011Proteasome}
\bibinfo{author}{van Ooijen, G.}, \bibinfo{author}{Dixon, L.~E.},
  \bibinfo{author}{Troein, C.} \& \bibinfo{author}{Millar, A.~J.}
\newblock \bibinfo{title}{Proteasome function is required for biological timing throughout the twenty-four hour cycle.}
\newblock \emph{\bibinfo{journal}{Current Biology}}
  \textbf{\bibinfo{volume}{21}}, \bibinfo{pages}{869--875}
  (\bibinfo{year}{2011}).
\newblock \urlprefix\url{http://dx.doi.org/10.1016/j.cub.2011.03.060}.

\bibitem{Yagita2010Development}
\bibinfo{author}{Yagita, K.} \emph{et~al.}
\newblock \bibinfo{title}{Development of the circadian oscillator during
  differentiation of mouse embryonic stem cells in vitro}.
\newblock \emph{\bibinfo{journal}{Proceedings of the National Academy of
  Sciences}} \textbf{\bibinfo{volume}{107}}, \bibinfo{pages}{3846--3851}
  (\bibinfo{year}{2010}).
\newblock \urlprefix\url{http://dx.doi.org/10.1073/pnas.0913256107}.

\bibitem{Merrow2001Circadian}
\bibinfo{author}{Merrow, M.} \emph{et~al.}
\newblock \bibinfo{title}{Circadian regulation of the light input pathway in
    \emph{Neurospora crassa}}.
\newblock \emph{\bibinfo{journal}{The EMBO Journal}}
  \textbf{\bibinfo{volume}{20}}, \bibinfo{pages}{307--315}
  (\bibinfo{year}{2001}).
\newblock \urlprefix\url{http://dx.doi.org/10.1093/emboj/20.3.307}.

\bibitem{Meng2008Setting}
\bibinfo{author}{Meng, Q.-J.} \emph{et~al.}
\newblock \bibinfo{title}{Setting clock speed in mammals: The CK1$\epsilon$ \emph{tau} mutation in mice accelerates circadian pacemakers by selectively destabilizing PERIOD proteins}.
\newblock \emph{\bibinfo{journal}{Neuron}} \textbf{\bibinfo{volume}{58}},
  \bibinfo{pages}{78--88} (\bibinfo{year}{2008}).
\newblock \urlprefix\url{http://dx.doi.org/10.1016/j.neuron.2008.01.019}.

\bibitem{Yin2006Nuclear}
\bibinfo{author}{Yin, L.}, \bibinfo{author}{Wang, J.}, \bibinfo{author}{Klein,
  P.~S.} \& \bibinfo{author}{Lazar, M.~A.}
\newblock \bibinfo{title}{Nuclear receptor Rev-erb$\alpha$ is a critical
  lithium-sensitive component of the circadian clock}.
\newblock \emph{\bibinfo{journal}{Science}} \textbf{\bibinfo{volume}{311}},
  \bibinfo{pages}{1002--1005} (\bibinfo{year}{2006}).
\newblock \urlprefix\url{http://dx.doi.org/10.1126/science.1121613}.

\bibitem{Iitaka2005Role}
\bibinfo{author}{Iitaka, C.}, \bibinfo{author}{Miyazaki, K.},
  \bibinfo{author}{Akaike, T.} \& \bibinfo{author}{Ishida, N.}
\newblock \bibinfo{title}{A role for glycogen synthase kinase-3$\beta$ in the mammalian circadian clock}.
\newblock \emph{\bibinfo{journal}{Journal of Biological Chemistry}}
  \textbf{\bibinfo{volume}{280}}, \bibinfo{pages}{29397--29402}
  (\bibinfo{year}{2005}).
\newblock \urlprefix\url{http://dx.doi.org/10.1074/jbc.m503526200}.

\bibitem{Besing2015Circadian}
\bibinfo{author}{Besing, R.~C.} \emph{et~al.}
\newblock \bibinfo{title}{Circadian rhythmicity of active GSK3 isoforms
  modulates molecular clock gene rhythms in the suprachiasmatic nucleus}.
\newblock \emph{\bibinfo{journal}{Journal of Biological Rhythms}}
  \textbf{\bibinfo{volume}{30}}, \bibinfo{pages}{155--160}
  (\bibinfo{year}{2015}).
\newblock \urlprefix\url{http://dx.doi.org/10.1177/0748730415573167}.

\bibitem{Narumi2016Mass}
\bibinfo{author}{Narumi, R.} \emph{et~al.}
\newblock \bibinfo{title}{Mass spectrometry-based absolute quantification
  reveals rhythmic variation of mouse circadian clock proteins}.
\newblock \emph{\bibinfo{journal}{Proceedings of the National Academy of
  Sciences}} \textbf{\bibinfo{volume}{113}}, \bibinfo{pages}{E3461--E3467}
  (\bibinfo{year}{2016}).
\newblock \urlprefix\url{http://dx.doi.org/10.1073/pnas.1603799113}.

\bibitem{Chen2015Emergent}
\bibinfo{author}{Chen, Y.}, \bibinfo{author}{Kim, J.~K.},
  \bibinfo{author}{Hirning, A.~J.}, \bibinfo{author}{Josi\'{c}, K.} \&
  \bibinfo{author}{Bennett, M.~R.}
\newblock \bibinfo{title}{Emergent genetic oscillations in a synthetic
  microbial consortium}.
\newblock \emph{\bibinfo{journal}{Science}} \textbf{\bibinfo{volume}{349}},
  \bibinfo{pages}{986--989} (\bibinfo{year}{2015}).
\newblock \urlprefix\url{http://dx.doi.org/10.1126/science.aaa3794}.

\bibitem{Tyson2015Models}
\bibinfo{author}{Tyson, J.~J.} \& \bibinfo{author}{Nov\'{a}k, B.}
\newblock \bibinfo{title}{Models in biology: lessons from modeling regulation of the eukaryotic cell cycle}.
\newblock \emph{\bibinfo{journal}{BMC Biology}} \textbf{\bibinfo{volume}{13}},
  \bibinfo{pages}{46} (\bibinfo{year}{2015}).
\newblock \urlprefix\url{http://dx.doi.org/10.1186/s12915-015-0158-9}.

\bibitem{OKeeffe2017Oscillators}
\bibinfo{author}{O'Keeffe, K.~P.}, \bibinfo{author}{Hong, H.} \&
  \bibinfo{author}{Strogatz, S.~H.}
\newblock \bibinfo{title}{Oscillators that sync and swarm}.
\newblock \emph{\bibinfo{journal}{Nature Communications}}
  \textbf{\bibinfo{volume}{8}}, \bibinfo{pages}{1504} (\bibinfo{year}{2017}).
\newblock \urlprefix\url{http://dx.doi.org/10.1038/s41467-017-01190-3}.

\end{thebibliography}
\end{document}